\documentclass{biometrika}

\usepackage{amsmath}

\usepackage{amsmath,amssymb}
\usepackage{booktabs}
\usepackage{xcolor}
\usepackage{iftex}
\usepackage{multirow}
\usepackage{newtxtext}
\usepackage[subscriptcorrection]{newtxmath}
\usepackage{etoolbox}
\usepackage{natbib}
\providecommand{\bibpreamble}{}
\usepackage{hyperref} 
\newcounter{coro}
\renewcommand{\thecoro}{\arabic{coro}}

\newenvironment{coro}{%
  \refstepcounter{coro}%
  \par\medskip
  \noindent{\normalfont\itshape Corollary~\thecoro.}\enspace
}{%
  \par\medskip
}

\usepackage{caption}

\graphicspath{{./art/}}

\usepackage[plain,noend]{algorithm2e}

\makeatletter
\renewcommand{\algocf@captiontext}[2]{#1\algocf@typo. \AlCapFnt{}#2} 
\def\@algocf@capt@plain{top}
\renewcommand{\algocf@makecaption}[2]{%
  \addtolength{\hsize}{\algomargin}%
  \sbox\@tempboxa{\algocf@captiontext{#1}{#2}}%
  \ifdim\wd\@tempboxa >\hsize
    \hskip .5\algomargin%
    \parbox[t]{\hsize}{\algocf@captiontext{#1}{#2}}
  \else%
    \global\@minipagefalse%
    \hbox to\hsize{\box\@tempboxa}
  \fi%
  \addtolength{\hsize}{-\algomargin}%
}
\makeatother

\makeatletter
\def\ps@plain{%
  \firstpage{\thepage}\let\@mkboth\@gobbletwo
  \def\@oddfoot{\hfil\thepage\hfil}%
  \let\@evenfoot\@oddfoot
  \def\@oddhead{}%
  \let\@evenhead\@oddhead
}
\makeatother

\def\S{\mathbf{S}}

\def\H{\mathbf{H}}

\def\A{\mathbf{A}}
\def\B{\mathbf{B}}
\def\C{\mathbf{C}}

\def\bR{\mathbf{R}}

\def\indi{\mathbb{I}}

\def\cG{\mathcal{G}}

\def\D{\mathbf{D}}

\def \R {\mathbb{R}}

\def \E {\mathbb{E}}

\newcommand{\Var}{\textup{Var}}
\newcommand{\Cov}{\textup{Cov}}
\newcommand{\Corr}{\textup{Corr}}

\begin{document}



\title{Generalized Independence Test}

\author{Mingshuo Liu\footnotemark[1]\thanks{* Liu and Zhou contributed equally to this work.}}
\affil{Department of Statistics, University of California, Davis}

\author{Doudou Zhou\footnotemark[1]}
\affil{Department of Statistics and Data Science, National University of Singapore}

\author{Hao Chen}
\affil{Department of Statistics, University of California, Davis}

\maketitle

\begin{abstract}
Testing independence is a widely encountered problem in modern data analysis, especially in high-dimensional settings where complex dependency structures are common. Traditional methods often lack power in such scenarios. To address this, we propose a novel test statistic that leverages both similarity and dissimilarity information to capture intricate relationships within the data. The proposed test demonstrates strong power across a wide range of alternatives in high-dimensional settings, as shown through extensive simulation studies. Under mild conditions, we prove that the permutation null distribution of the statistic converges to the $\chi^2_4$ distribution, enabling straightforward control of the type I error. Our theoretical analysis further advances the moment method to establish joint asymptotic normality for a class of double-indexed permutation statistics.  Moreover, we prove the power consistency of the proposed test without imposing explicit restrictions on the data dimension, under mild conditions that allow for detecting complex dependencies. We illustrate the proposed test using gene-expression data from the Genotype-Tissue Expression project, where GIT rejects independence between the GTEx category ``Breast - Mammary Tissue'' and the cell-line category ``Cells - EBV-transformed lymphocytes.''
\end{abstract}


\section{Introduction }
\label{Section: Introduction}
Testing the independence between variables is a fundamental question across diverse research areas. For instance, \cite{wilks1935independence} used the likelihood ratio statistic to test independence for several sets of normally distributed random variables; \cite{feige1979casual} explored the independence relationships between economic variables such as money supply and nominal income; \cite{martin2005testing} investigated quasi-independence between failure and truncation times in survival analysis; and \cite{liu2010versatile} examined gene associations in genome-wide association studies.

Given two random variables, $X$ and $Y$, with marginal distributions $P_X$ and $P_Y$ over the spaces $\mathcal{X}$ and $\mathcal{Y}$, respectively, and their joint distribution $P_{XY}$ on $\mathcal{X} \times \mathcal{Y}$, the goal is to test the hypothesis:
\begin{align*}
H_0: P_{XY}=P_X P_Y \quad \text{ versus } \quad H_1: P_{XY} \neq P_X P_Y    
\end{align*}
using paired samples $\{ (X_i,Y_i) \}_{i=1}^n$  drawn independently and identically from $P_{XY}$.

For univariate data, where $X \in \R$ and $Y \in \R$, classical methods such as Pearson correlation \citep{pearson1895vii}, Spearman's $\rho$ \citep{spearman1904proof}, and Kendall's $\tau$ \citep{kendall1938new} provide traditional means of analysis. These test statistics can be expressed in the following generalized coefficient form (before standardization):
\begin{equation}
 \Gamma  = \sum_{i=1}^n \sum_{j=1}^n A_{ij} B_{ij} \,.
\label{equa: def Gamma}
\end{equation}
Here, $A_{ij}$ and $B_{ij}$ represent measures of certain  properties between observations $(X_i, X_j)$ and $(Y_i, Y_j)$, respectively. The specific configurations: $A_{ij}=X_i - X_j, B_{ij}=Y_i - Y_j$; $A_{ij}=\text{rank}(X_i) -\text{rank}(X_j) , B_{ij}=\text{rank}(Y_i) - \text{rank}(Y_j)$; $ A_{ij}=\text{sign}(X_i-X_j) , B_{ij}=\text{sign}(Y_i-Y_j)$,
correspond to the Pearson correlation, Spearman's $\rho$, and  Kendall's $\tau$ measures, respectively. However, these methods are not effective for detecting non-monotonic dependency structures, prompting the development of various alternative measures of statistical association \citep{hirschfeld1935connection, hoeffding1994non, blum1961distribution}. Recently, the binary expansion technique has been utilized by \cite{zhang2019bet} for testing independence, and {  \cite{chatterjee2021new} introduced a simple rank-based estimator of a population dependence
measure previously studied by \cite{dette2013copula} and \cite{gamboa2018sensitivity}, which~is
$0$ if and only if the variables are independent and $1$ if and only if $Y$ is a measurable function of~$X$.}

Nowadays, data analysis faces significant challenges due to the advent of multivariate and high-dimensional data. There has been considerable interest in testing independence in these complex datasets. For example, in the Gait data, researchers examined the relationship between the angular rotations of the hip and knee, and in the Canadian Weather data, they investigated the relationship between the time series of temperature and precipitation \citep{sarkar2018some}. Additionally, in gene expression (RNA-seq) and chromatin accessibility (ATAC-seq) studies, testing independence is crucial because RNA-seq and ATAC-seq are widely assumed to co-vary, and their dimensions reach tens of thousands \citep{cai2023asymptotic}. For multivariate or high-dimensional data, parametric models are typically tailored to specific distributions and often struggle to accommodate data from diverse contexts.

In the nonparametric domain, \cite{friedman1983graph} utilized similarity graphs constructed from pairwise distances among observations. Their test statistics can also be expressed in the form of \eqref{equa: def Gamma}. Subsequently, \cite{szekely2007measuring} and \cite{gretton2007kernel, zhang2024fast} approached the problem by defining $A_{ij}$ and $B_{ij}$ through centered pairwise distances and kernel values, respectively. The latter two methods leveraged more similarity information than the method in \cite{friedman1983graph}.

Rank-based methods are notable for their robustness and simplicity. Several rank-based tests have been developed for testing independence. \cite{heller2013consistent} used the ranks of pairwise distances between the sample values of $X$ and of $Y$ to construct the test statistic by constructing contingency tables for all sample points. \cite{moon2022interpoint} generalized the sign covariance introduced by \cite{bergsma2014consistent}, extending Kendall's $\tau$ by incorporating interpoint ranking into a U-statistic. \cite{shi2022distribution} combined distance covariance \citep{szekely2007measuring} with the center-outward ranks and signs developed by \cite{hallin2017distribution}, resulting in a consistent and distribution-free test in the family of multivariate distributions with non-vanishing Lebesgue probability densities. \cite{deb2023multivariate}  leveraged multivariate ranks derived from measure transport theory, achieving a test statistic that remains distribution-free with respect to $P_X$ and $P_Y$ under the null hypothesis.  A common limitation across these methodologies is the absence of practical asymptotic formulas. Specifically, determining critical values in~\cite{deb2023multivariate} requires substantial sample sizes to establish thresholds, while other approaches rely on permutation testing or estimating the quantile of a quadratic form, which can be computationally~intensive.

Beyond rank-based approaches, several nonparametric tests have been developed based on interpoint distances. For example, \cite{biswas2016some,sarkar2018some,guo2020nonparametric} directly used distance information between sample points in constructing their test statistics, avoiding the use of ranks. Other nonparametric methods for multivariate or high-dimensional independence testing include those based on mutual information \citep{berrett2019}, binary expansion \citep{zhang2024beauty}, and deep neural networks \citep{cai2023asymptotic}.

Despite the progress in this field, existing methods still exhibit limitations when applied to certain straightforward scenarios.  Consider  $p$-dimensional paired samples $\{ (X_i,Y_i) \}_{i=1}^n$, where the dependency structure is defined as $Y_{ij}=\log(|X_{ij}|)$, $i=1,\ldots,n$, $j=1,\ldots,p$. Here, $n=150$ represents the sample size, and $p=50$ represents the dimension. The values of $X_{ij}$ are generated independently from $N(0,1)$. { For this illustrative example, we estimate the empirical power at the nominal
level $0.05$.}  We perform a comparison of ten existing independence tests.
These tests cover a range of approaches, including: graph-based methods \citep{friedman1983graph} (FR1 and FR2), distance correlation using the R package \textit{energy} \citep{szekely2007measuring} (dCov), rank of distances using R package \textit{HHG} \citep{heller2013consistent} (HHG), projection correlation \citep{zhu2017projection} (pCov), mutual information using R package \textit{IndepTest} \citep{berrett2019} (MINT), interpoint-ranking sign covariance \citep{moon2022interpoint} (IPR), a test by combining distance covariance with the center-outward ranks and signs \citep{shi2022distribution} (Hallin), a multivariate rank-based test using measure transport \citep{deb2023multivariate} (RdCov), and a circularly permuted classification based test \citep{cai2023asymptotic} (CPC). The tuning parameters for these comparative methods are set to their default values.

The results are shown in Table \ref{tab: motivate eg2}. We observe that, except for FR1 and FR2, which have power values of 0.486 and 0.521, respectively, all other methods have power less than 0.2. Even for FR2, the power of 0.521 is not particularly high.

\begin{table}[!htbp]
\caption{The estimated power for $Y_{ij}=\log(|X_{ij}|)$ based on 1000 replications.}
\centering
{\small
\begin{tabular}[!htbp]{l|c|c|c|c|c|c|c|c|c|c|c}
\hline
Method & New & FR1 & FR2 & dCov & HHG &  pCov & MINT & IPR & Hallin & RdCov & CPC \\
\hline
Power & $\mathbf{0.952}$ & 0.486 & 0.521 & 0.178 & 0.130 & 0.135 & 0 & 0.123 & 0.061 & 0.050 & 0.067 \\
\hline
\end{tabular}
}
\label{tab: motivate eg2}
\end{table}

To address these limitations, we introduce the \textbf{Generalized Independence Test (GIT)}, a novel methodology designed for high-dimensional  data contexts. Although GIT is not limited to rank-based measures, in this paper we illustrate its application using ranks derived from similarity and dissimilarity graphs. As shown in Table \ref{tab: motivate eg2}, the default version of our method, GIT-$\text{R}_{\text{r}k\text{NF}}$ (denoted as `New' in the table), which leverages \textbf{R}anks from a \textbf{r}obust $k$-\textbf{N}earest neighbor graph and a robust $k$-\textbf{F}arthest point graph introduced in Section \ref{sec: ranks and test stat}, has a power of 0.952 in the above example.
{ The limiting null theory below is developed at the level of general similarity and dissimilarity matrices. {For the graph-specific order verification, we cover both ordinary and robust $k$-NNG-based constructions in fixed dimension, where the needed degree and row-sum bounds can be handled analytically. For the power analysis, we work with the canonical constructions in Examples 1 and 4, and provide a result for Example 5 under a replacement-stability condition.} In the numerical studies, we use the robust $k$-NNG and robust $k$-farthest point graph \citep{zhu2023new} for { better small-sample performance}, and { Supplement S9 reports numerical diagnostics for the remaining conditions for the full robust nearest/farthest construction, including the cross-component conditions required for joint normality.}}
Our main contributions can be summarized as follows. 
First, we propose an independence test that combines graph-based similarity and dissimilarity information in a unified statistic.  
By aggregating these two complementary views, the procedure achieves high power against a wide range of complex dependence structures, as illustrated in Section~\ref{Section: Simulation studies}.  
We also show that, under mild conditions on the underlying graphs, the test is power consistent without imposing explicit constraints on the data dimension. Second, we prove a univariate combinatorial central limit theorem for the proposed class of double-indexed statistics under several alternative sets of assumptions, each of which is sufficient on its own.  
These conditions allow us to cover both sparse and dense graph regimes when the statistic is graph-based, whereas existing results typically require very dense graphs. Third, we establish a multivariate combinatorial central limit theorem for vectors of double-indexed statistics, extending the classical result of \citet{daniels1944relation} beyond the single-statistic setting.  
Our assumptions again accommodate a broad range of graph sparsity levels and yield a flexible framework for studying the joint asymptotic behavior of multiple graph-derived functionals. Finally, we show that the proposed method is asymptotically distribution-free and does not rely on moment assumptions. { Under mild regularity conditions, the permutation null distribution converges to a $\chi^2_4$ limit, which provides analytic critical values without the need for additional resampling. The simulations in Section \ref{Section: Simulation studies} show that this analytic calibration yields reasonably well-controlled empirical size in the small-sample, high-dimensional settings considered.}

The rest of the paper is organized as follows. The details of the new test are presented in Section~\ref{sec: ranks and test stat}. In Section~\ref{Section: Asymptotic properties}, we prove that the proposed test statistic  is asymptotic distribution-free, enabling straightforward type-I error control. {We also provide the power consistency results under mild conditions.} Extensive performance analysis in Section \ref{Section: Simulation studies} shows that GIT is particularly powerful at capturing complex dependency structures.  In Section \ref{Section: Real data analysis}, we  apply GIT to Genotype-Tissue Expression data, and conclude with discussions in Section \ref{sec: discussion}. The proofs of
the theorems and corollaries are provided in the Supplement Material. {
Code implementing our method and the real data used in Section~\ref{Section: Real data analysis} are available at \url{https://github.com/mingshuostat/generalized-independence-test}.}

\section{Generalized independence test}
\label{sec: ranks and test stat}

We begin by introducing the general formulation of our test statistic, which is derived from sample similarity and dissimilarity matrices. Let $\mathbf{S}^Z = [S^Z_{ij}]_{i,j=1}^n \in  \R^{n \times n}$ and $\mathbf{D}^Z = [D^Z_{ij}]_{i,j=1}^n \in  \R^{n \times n}$ represent the similarity and dissimilarity matrices for observations in sample $Z$, where $Z \in \{X, Y\}$. Optional measures of similarity and dissimilarity will be outlined later in this section. From these matrices, we define four generalized correlations as follows:
\begin{align*}
T_1 =\sum_{i=1}^n \sum_{j\neq i}^n D^X_{ij} D^Y_{ij}, \,
T_2=\sum_{i=1}^n \sum_{j\neq i}^n D^X_{ij} S^Y_{ij},  \, T_3 =\sum_{i=1}^n \sum_{j\neq i}^n S^X_{ij} D^Y_{ij}, \,
T_4=\sum_{i=1}^n \sum_{j\neq i}^n S^X_{ij} S^Y_{ij} \,.
\end{align*}
These correlations are designed to effectively capture various types of dependency between $X$ and $Y$, acknowledging the complexity of real-world dependency. To address these diverse scenarios, we propose the Generalized Independence Test (GIT), formulated as:
%
%
\begin{equation}
T = (T_1-\mu_1, T_2-\mu_2, T_3-\mu_3, T_4-\mu_4) \mathbf \Sigma^{-1} (T_1-\mu_1, T_2-\mu_2, T_3-\mu_3, T_4-\mu_4)^{\top},
\label{eqn: test stat2}
\end{equation}
%
%
where $\mu_s=\E(T_s)$ for $s \in \{1,2,3,4\}$, and $\mathbf \Sigma=\Cov\big((T_1, T_2, T_3, T_4)^{\top}\big)$. Here, $\E$ and $\Cov$ denote the expectation and covariance matrix under the permutation null distribution, respectively, calculated for the dataset 
$\{ (X_i,Y_{\pi(i)}) \}_{i=1}^n$, where $\pi$ denotes a permutation of $\{1,\ldots,n\}$ with each permutation equally probable.

By constructing $T$ in this way, we aim to capture the intricate dependency structure that arises when the patterns of similarity and dissimilarity in $X$ are related to those in $Y$. For instance, similarity in $X$ might correspond to  dissimilarity in $Y$, which would be reflected in $T_3$, or to similarity in $Y$, which would correspond to $T_4$. Similarly, $T_1$ and $T_2$ capture other types of relationships between $X$ and $Y$. This formulation of $T$ ensures that any deviation of the four generalized correlations from their expected values under the null hypothesis of independence contributes to the overall test statistic, potentially resulting in a large value for $T$. We will reject the null hypothesis if $T$ exceeds the critical value for a specified nominal level.

Our test statistic offers a versatile framework, allowing flexibility in the choice of similarity and dissimilarity matrices, including those based on unweighted graphs and weighted graphs. Following are some examples.

\noindent\emph{\textbf{Example 1} (Unweighted Graph)}. Similarity and dissimilarity graphs, denoted as $G_S^Z$ and $G_D^Z$, can be used to define the similarity and dissimilarity matrices as follows:
$$
S_{ij}^Z = 
\begin{cases}
    1, & \text{if } (i,j) \in G_S^Z, \\
    0, & \text{otherwise},
\end{cases}
\text{ and }~ 
D_{ij}^Z = 
\begin{cases}
    1, & \text{if } (i,j) \in G_D^Z, \\
    0, & \text{otherwise}.
\end{cases}
$$
Several options are available for constructing similarity graphs, such as the $k$-nearest neighbor graph ($k$-NNG) and the $k$-minimum spanning tree ($k$-MST)\footnote{The MST is a spanning tree connecting all observations while minimizing the total edge length. The $k$-MST is the union of the first through $k$th MSTs, where the $k$th MST minimizes the total edge length while excluding edges from previous MSTs.} \citep{friedman1979multivariate}. For dissimilarity graphs, alternatives include the $k$-farthest point graph, where each node connects to its $k$ farthest points, and the $k$-maximal spanning tree, which mirrors the construction of the $k$-MST but maximizes total edge lengths. These graphs follow the same principles as similarity graphs, but with objectives based on dissimilarity. For example, in the $k$-NNG, each $Z_i$ is linked to the $k$ observations minimizing $\|Z_i - Z_j\|$, where $\|\cdot\|$ indicates the Euclidean norm; in contrast, the $k$-farthest point graph links to those maximizing this distance.

\noindent\emph{\textbf{Example 2} (Weighted Graph: Pairwise Distances).} After constructing the similarity and dissimilarity graphs, we assign weights based on pairwise distances. For example, the weights can be defined as follows:
$$
S_{ij}^Z = 
\begin{cases}
    -\|Z_i - Z_j\|, & \text{if } (i,j) \in G_S^Z, \\
    0, & \text{otherwise},
\end{cases}
\text{ and } ~
D_{ij}^Z = 
\begin{cases}
    \|Z_i - Z_j\|, & \text{if } (i,j) \in G_D^Z, \\
    0, & \text{otherwise}.
\end{cases}
$$

\noindent\emph{\textbf{Example 3} (Weighted Graph: Kernels).} Similar to Example 2, but using kernel values as weights. For instance, the weights can be defined with appropriate choices of $\sigma_{Z S}^2$ and $\sigma_{Z D}^2$:
$$
S_{ij}^Z  \!=  \!
\begin{cases}
 \exp \!\big(\! - \! \frac{\|Z_i - Z_j\|^2}{2 \sigma_{Z S}^2} \big)\!, & \!\text{if } (i,j) \in G_S^Z, \\
    0, & \text{otherwise},
\end{cases}
~ 
D_{ij}^Z \!= \!
\begin{cases}
     \! -\! \exp\! \big(\! - \!\frac{\|Z_i - Z_j\|^2}{2 \sigma_{Z D}^2} \big)\!, & \!\text{if } (i,j) \in G_D^Z, \\
    0, & \text{otherwise}.
\end{cases}
$$

\noindent\emph{\textbf{Example 4} (Weighted Graph: Graph-induced Ranks).} The matrices $\S^Z$ and $\D^Z$ are constructed using the graph-induced rank proposed by \citet{zhou2023new}. For two graphs $G^1$ and $G^2$ sharing the same vertices, their union, $G^1 \cup G^2$, combines all edges from both graphs. If they do not have shared edges, $G^1 \cap G^2 = \emptyset$. Starting with an edgeless graph $G^0$, we iteratively build a sequence of graphs $\{ G^l \}_{l=0}^k$ as follows:
$
G^{l+1}  = G^{l} \cup \widetilde G^{l+1}, ~ \widetilde G^{l+1} = \arg \max_{G' \in \cG^{l+1} } \sum_{(i,j) \in G'} S(Z_i,Z_j), 
$
where $\cG^{l+1}$ contains graphs disjoint from $G^l$, i.e., $\cG^{l+1} = \{G' \in \cG: G'\cap G^{l} =  \emptyset\}$, and $\cG$ is a graph set whose elements satisfy specific user-defined constraints. $S(\cdot,\cdot)$ is a similarity measure, such as  $S(Z_i,Z_j) = -\|Z_i - Z_j\|$. This method can be used to construct common similarity graphs, like $k$-NNG and $k$-MST. The graph-induced rank matrix $\widetilde {\mathbf{R}} =[\widetilde R_{ij}]_{i,j=1}^n$ is then defined as $ \widetilde R_{i j } = \sum_{l=1}^k \indi\big( (i,j) \in G^l \big), \quad 1 \leq i,j \leq n$,
where $\indi(A)$ is $1$ if the event $A$ occurs and $0$ otherwise. This rank assigns greater weights to edges with higher similarity, enriching the unweighted graph with more detailed similarity information. For instance, an edge in the $l$th-NNG of a $k$-NNG has a rank of $k - l + 1$. This rank-based approach improves robustness by reducing sensitivity to outliers compared to direct distance measurements.

While initially defined for similarity graphs, the rank can also be extended to dissimilarity matrices using dissimilarity graphs, such as the $k$-farthest point graph or $k$-maximal spanning tree. Thus, $\mathbf{S}^Z$ and $\mathbf{D}^Z$ represent the graph-induced rank matrices for similarity and dissimilarity graphs, respectively.

\noindent\emph{\textbf{Example 5} (Weighted Graph: Ranks Based on Robust Graphs).} This example introduces ranks based on the robust $k$-Nearest Neighbor Graph (r$k$-NNG) proposed by \cite{zhu2023new}. Given a distance metric $D(\cdot, \cdot)$, let $R_i(Z_j)$ denote the rank of the distance $D(Z_i, Z_j)$ among the set of distances $\{D(Z_i, Z_l)\!: \! l \neq i \}$. Define $|G_i|$ as the degree of $i$-th node in a graph $G$ constructed on observations $\{Z_i\}_{i=1}^n$, and let $O_i$ be the set of $k$ observations excluding $Z_i$. The r$k$-NNG is the graph that minimizes $
\sum_{i=1}^n \sum_{x \in O_i} R_{i}(x) + \lambda \sum_{i=1}^n |G_i|^2$, 
where $\lambda$ is a hyper-parameter that penalizes hubs (nodes with high degree). When $\lambda=0$, this reduces to the standard $k$-NNG. With $\widetilde G$ representing the r$k$-NNG, {we define the rank matrix based on $\widetilde G$ as $\widetilde {\mathbf{R}} =[\widetilde R_{ij}]_{i,j=1}^n$}, where $\widetilde R_{ij}$ represents the rank of the distance $D(Z_i, Z_j)$ among all observations connected to $Z_i$, i.e.,
\begin{align*}
\widetilde R_{ij} = \begin{cases}
    \sum_{x \in O_i} \indi(D(Z_i, Z_j) \leq D(Z_i, x)), & \text{if } (i,j) \in \widetilde G, \\
    0, & \text{otherwise}.
\end{cases} .
\end{align*}

Then, $\S^Z$ can be defined as $\widetilde {\mathbf{R}}$. Similarly, $\D^Z$ can be defined in the same way, but using the robust $k$-farthest point graph, where the distance $D(Z_i,Z_j)$ used in constructing the r$k$-NNG is replaced by $ - D(Z_i,Z_j)$.
The choices of $\S^Z$ and $\D^Z$  are not limited to the examples above; many other options are available. Our focus is not on identifying the best choice but rather on providing a general framework of \eqref{eqn: test stat2}. We will illustrate  performance using ranks based on r$k$-NNG and robust $k$-farthest point graph (Example 5) in the following due to their robustness and ability to capture extensive information. For the $k$-NNG and $k$-farthest point graph constructions in Examples 4 and 5, the initial rowwise rank matrices are not necessarily symmetric. We follow \cite{zhou2023new} to symmetrize them by replacing $\widetilde {\mathbf{R}}$ with $\frac{1}{2}(\widetilde {\mathbf{R}} + \widetilde {\mathbf{R}}^{\top})$. We assume that the pairwise distances within each sample are almost surely distinct, so that each row of the initial rank matrix contains the nonzero ranks $1,\ldots,k$. This symmetrization yields comparable test performance and simplifies the theoretical analysis. In the following, we refer to the Generalized Independence Test \eqref{eqn: test stat2} with symmetrized $\S^Z$ and $\D^Z$ defined in Example 5 as GIT-$\text{R}_{\text{r}k\text{NF}}$.  Further exploration of potentially better choices may be pursued in future work. The hyper-parameter $\lambda$ in Example 5 is set to $0.3$, following the recommendation by \cite{zhu2023new}, and $k$ is set to be $\lfloor n^{0.5} \rfloor$, which will be explored in Supplement S7.

\section{Theoretical properties}
\label{Section: Asymptotic properties}
In this section, we study the asymptotic properties of the test statistic, which are crucial for deriving an analytic approximation of the $p$-value. Recall the test statistic
\begin{equation*}
T = (T_1-\mu_1, T_2-\mu_2, T_3-\mu_3, T_4-\mu_4) \mathbf \Sigma^{-1} (T_1-\mu_1, T_2-\mu_2, T_3-\mu_3, T_4-\mu_4)^{\top}.
\end{equation*}
We begin by analyzing $T_s, s=1,2,3,4$, which can be written in the form:
$$
T_s = \sum_{i=1}^n \sum_{j = 1}^n A_{ij}^{(s)} B_{ij}^{(s)},
$$
where $\A^{(1)}=\A^{(2)}=\mathbf{D}^X$, $\A^{(3)}=\A^{(4)}=\mathbf{S}^X$, $\B^{(1)}=\B^{(3)}=\mathbf{D}^Y$ and $\B^{(2)}=\B^{(4)}=\mathbf{S}^Y$, with all matrices symmetric and having zero as diagonals. We study their limiting distribution under the permutation null distribution, based on the permuted data 
$\{ (X_i,Y_{\pi(i)}) \}_{i=1}^n$, where $\pi$ denotes a permutation of $\{1,\ldots,n\}$ with each permutation equally probable.

\subsection{Moment properties}
\label{subsec: moment property}

We first derive the analytic expressions for
$\E(T_s)$ and $\Cov(T_s, T_{s'}),$ $s, s' \in \{1,2,3,4\}$ through combinatorial analysis.
To simplify the notations, for $\C^{(s)} \in \{\A^{(s)}, \B^{(s)}\}$, we denote
\begin{equation}
\label{eqn:C123,def}
\begin{aligned}
& C_{1}^{(s)} = \sum_{i=1}^{n} \sum_{j=1}^n  C_{ij}^{(s)}, \quad C_{i \cdot}^{(s)} = \sum_{j=1}^n C_{ij}^{(s)}, \text{ for } i \in \{1,\ldots,n\},   \\
&  C_2^{(ss')} = \sum_{i=1}^{n} \sum_{j=1}^n  C_{ij}^{(s)} C_{ij}^{(s')}, \quad  C_3^{(ss')}= \sum_{i=1}^{n} C_{i \cdot}^{(s)} C_{i \cdot}^{(s')}.
\end{aligned}    
\end{equation}

\begin{theorem}
\label{thm: expect, cov}
  {For $n\geq4$,} under the permutation null distribution, we have
$\E(T_s)  = \frac{A_1^{(s)} B_1^{(s)}}{n(n-1)}$ and $\Cov(T_s,T_{s'})$ is 
\begin{align*}
& \frac{4(n+1) (A_3^{(ss')} \!-\!\frac{A_1^{(s)} A_1^{(s')}}{n}) (B_3^{(ss')}\!-\!\frac{B_1^{(s)} B_1^{(s')}}{n}) }{n(n-1)(n-2)(n-3)}  + \frac{2(A_2^{(ss')}\!-\!\frac{A_1^{(s)} A_1^{(s')}}{n(n-1)}) (B_2^{(ss')}\!-\!\frac{B_1^{(s)} B_1^{(s')}}{n(n-1)})}{n(n-3)} \\
&  - \frac{4(A_2^{(ss')}\!-\!\frac{A_1^{(s)} A_1^{(s')}}{n(n-1)})(B_3^{(ss')}\!-\!\frac{B_1^{(s)} B_1^{(s')}}{n})}{n(n-2)(n-3)}  - \frac{4(A_3^{(ss')}\!-\!\frac{A_1^{(s) }A_1^{(s')}}{n})(B_2^{(ss')}\!-\!\frac{ B_1^{(s)} B_1^{(s')}}{n(n-1)})}{n(n-2)(n-3)}. 
\end{align*}
\end{theorem}
The proof of Theorem \ref{thm: expect, cov} is presented in Supplement S1.
For the test statistic $T$ to be well-defined, it is crucial that the covariance matrix $\mathbf{\Sigma}$ is positive-definite. This condition is not restrictive and it is feasible to verify the finite-sample positive-definiteness of $\mathbf{\Sigma}$. If ${\rm rank}(\mathbf{\Sigma}) < 4$, it implies that the statistics $T_s$, for $s \in \{ 1,2,3,4\}$, are linearly dependent. In such cases, using only the linearly independent statistics is sufficient to construct the test statistic. In the following, we will explore the asymptotic distribution of $T_s$ under the permutation null distribution and examine the conditions under which $\Sigma$ is asymptotically positive-definite.

\subsection{Asymptotic properties}
\label{subsec: asymp}

{
We first introduce the following notations. For two sequences of nonnegative real numbers $\{a_n\}$ and $\{b_n\}$, we denote $a_{n}=o(b_{n})$ or $a_{n}\prec b_{n}$ if $a_{n}$ is dominated by $b_{n}$ asymptotically, i.e., $\lim _{n \rightarrow \infty} a_n/b_n=0$, $a_{n} \asymp b_{n}$ if $a_{n}$ is bounded both above and below by $b_{n}$ up to a constant factor asymptotically, $a_{n}=O(b_{n})$ or 
$a_{n} \precsim b_{n}$ if  $a_{n}$ is bounded above by $b_{n}$ up to a constant factor asymptotically. 
}

The asymptotic normality of the generalized correlation coefficient, $\Gamma$ in \eqref{equa: def Gamma}, has been previously studied by \cite{daniels1944relation, pham1989asymptotic}. More recently, \cite{huang2023kernel} introduced a novel set of conditions  allowing for more flexible scaling relationships than those in \cite{pham1989asymptotic}. Notably, \cite{daniels1944relation} requires that $\A^{(s)}$ and $\B^{(s)}$ are anti-symmetric. However, the anti-symmetry is only needed to ensure $A_1=B_1=0$, and it can be replaced by the conditions $A_1=B_1=0$ along with $\A^{(s)}$ and $\B^{(s)}$ being symmetric \citep{friedman1983graph, zhu2021note}. To understand how these results apply to our statistics $T_s$ for $s=1,\dots, 4$, we review the conditions in these earlier works.
Let $G^X$ and $G^Y$ represent the similarity or dissimilarity graphs constructed from the $X$  and $Y$ samples, as described in Section \ref{sec: ranks and test stat}. We allow $k$ to vary with $n$ such that $k \asymp n^{\alpha}$ for some $\alpha \in [0,1]$. Specifically, for unweighted graphs, we consider the graphs in which each observation { is connected to at least $ck$ other observations for some constant $c>0$}, and $|G^Z|$ is $O(nk)$, where $|G^Z|$ is the number of edges in the graph $G^Z$ for $Z \in \{X,Y\}$. For weighted graphs, we focus on those described in Examples 4 and 5 ($k$-NNG, $k$-MST, r$k$-NNG, $k$-farthest point graph, $k$-maximal spanning tree, robust $k$-farthest point graph), with graph-induced ranks (Example 4) or ranks based on robust graph (Example 5). It is important to note that the conditions in \cite{daniels1944relation} hold only when $\alpha=1$, and the conditions in \citet[][Theorem~3.2]{pham1989asymptotic} only apply when $\alpha=0$.  However, the conditions in
\citet[][Theorem~C.1]{huang2023kernel} and \citet[][Theorem~3.1]{pham1989asymptotic} cannot be satisfied when $\A^{(s)}$ and $\B^{(s)}$ are derived from unweighted graphs or weighted graphs using graph-induced ranks and ranks based on robust graphs. Interestingly, as shown in Supplement S7, optimal performance of GIT-$\text{R}_{\text{r}k\text{NF}}$  often occurs when $\alpha \in [0.4, 0.7]$. Recognizing the gap between $\alpha=0$ and $\alpha=1$, and the practical need for using $\alpha \in (0,1)$, we propose a set of conditions to bridge this gap. Specifically, our conditions can potentially be satisfied when $\alpha \in (0,1]$.

To simplify the derivation, we consider centered versions of the { matrices $\C^{(s)} \in \{\A^{(s)}, \B^{(s)} \}$, $s \in \{1,2,3,4\}$}, such that $\sum_{i=1}^n \sum_{j=1}^n C_{ij}^{(s)} = 0$. If this does not hold, we will replace $C_{ij}^{(s)}$ by  $C_{ij}^{(s)} - \frac{C_{1}^{(s)}\indi(i \neq j)}{n(n-1)}$, and this centering will not change the value of the generalized test statistic $T$ in \eqref{eqn: test stat2}. Then these centered matrices satisfy the following Condition \ref{cond0} automatically.

\begin{condition}
$C_{ii}^{(s)} =  0, C_{i j}^{(s)} = C_{j i}^{(s)}, i,j \in \{1, \ldots, n\}$, $C^{(s)}_{1} = 0 $, for $\C^{(s)} \in \{\A^{(s)},$ $ \B^{(s)}\}$ and $s \in \{1,2,3,4\}$.   
\label{cond0}
\end{condition}
{For $s,s' \in \{1,2,3,4\}$ and $\C \in \{\A,\B\}$, define}
\begin{equation*}
\begin{aligned}
& C^{(s)}_{0+}=\max_{1\leq i,j \leq n} |C^{(s)}_{ij}|, ~ C_{i \star}^{(s)} = \sum_{j=1}^n |C_{ij}^{(s)}|, ~   C_{1+}^{(s)} = \max_{1 \leq i \leq n} C_{i \star}^{(s)},\\
& C_{2+}^{(ss')} = \sum_{i=1}^{n} \sum_{j=1}^n  |C_{ij}^{(s)} C_{ij}^{(s')}|, ~ C_{3+}^{(ss')}= \sum_{i=1}^{n}  C_{i \star}^{(s)} C_{i \star}^{(s')}.
\end{aligned}    
\end{equation*}
Notably, $C_{2+}^{(ss)} = C_2^{(ss)}$. We proceed to introduce the following Conditions \ref{Condition: Condition 1 in CLT normal}-\ref{Condition: Condition 3 in CLT normal}. Satisfying any one of these conditions is sufficient for the asymptotic normality of $T_s$ in Theorem \ref{Thm: CLT general form, one stat}. These conditions essentially capture different dominant terms in variances associated with each statistic, as shown in Corollary S1 in Supplement S2.

\begin{condition}
\label{Condition: Condition 1 in CLT normal}
{$\max\{ n^2 (A_{0+}^{(s)} B_{0+}^{(s)})^2,\!$  $(A_{1+}^{(s)} B_{1+}^{(s)})^2,\!$ $n^{-1} A_{3+}^{(ss)} B_{3+}^{(ss)}\} \! \prec\! A_{2+}^{(ss)} B_{2+}^{(ss)}$.}
\end{condition}
\begin{condition}
\label{Condition: Condition 2 in CLT normal} 
$\max\{ n^{3} (A_{0+}^{(s)} B_{0+}^{(s)})^2,$ $ n (A_{1+}^{(s)} B_{1+}^{(s)})^2,$ $n A_{2+}^{(ss)} B_{2+}^{(ss)} \} \prec A_{3+}^{(ss)} B_{3+}^{(ss)}$ $\asymp$ $A_{3}^{(ss)} B_{3}^{(ss)}$.
\end{condition}
\begin{condition}
\label{Condition: Condition 3 in CLT normal} 
{$C_{2+}^{(ss)} \prec C_{3+}^{(ss)}$ for $\C\in\{\A,\B\}$, $\max\{n^{3} (A_{0+}^{(s)} B_{0+}^{(s)})^2 , n (A_{1+}^{(s)} B_{1+}^{(s)})^2 \} \prec$ $n A_{2+}^{(ss)} B_{2+}^{(ss)}$ $\asymp A_{3+}^{(ss)} B_{3+}^{(ss)}$ $ \asymp A_{3}^{(ss)} B_{3}^{(ss)}$.}
\end{condition}
\begin{theorem}
\label{Thm: CLT general form, one stat}
{For any fixed $s\in\{1,2,3,4\}$}, assume that $\A^{(s)}$ and $\B^{(s)}$ satisfy Condition \ref{cond0}. If  $\A^{(s)}$ and $\B^{(s)}$ further satisfy one of Conditions \ref{Condition: Condition 1 in CLT normal}-\ref{Condition: Condition 3 in CLT normal}, under the permutation null distribution,
\begin{equation}
\frac{T_s }{ \sqrt{ {\rm Var}(T_s) }} \stackrel{\mathcal{D}}{\rightarrow}  N(0,1) ~ \text{ when } n \rightarrow \infty, 
\label{Equation: CLT convergence normal distribution}
\end{equation}
 where $\stackrel{\mathcal{D}}{\rightarrow}$ denotes convergence in distribution.
\end{theorem}

The proof of Theorem \ref{Thm: CLT general form, one stat} is provided in Supplement S2, where we show a case-wise analysis under Conditions \ref{Condition: Condition 1 in CLT normal}-\ref{Condition: Condition 3 in CLT normal}. This proof predominantly utilizes the technique of moments matching, a method that has been crucial in showing the asymptotic behavior for the double-indexed linear permutation statistics including 
 graph-based statistics \citep{henze1988multivariate,petrie2016graph,huang2023kernel} and kernel-based statistics \citep{song2023practical,song2023}.

\begin{remark}
\label{remark: cond table}
    To better understand Conditions \ref{Condition: Condition 1 in CLT normal}-\ref{Condition: Condition 3 in CLT normal}, we simplify the choice of graphs and restate the conditions under these simplified graphs in Tables \ref{tab: para relation1} and \ref{tab: para relation2}. In these tables, we omit the superscripts and use $\C$ to represent $\C^{(s)},$ for $\C^{(s)} \in \{\A^{(s)}, \B^{(s)}\}$. Specifically, for unweighted graphs in Table \ref{tab: para relation1}, we consider graphs where each observation { is connected to at least $ck$ other observations for some constant $c>0$}, and $|G^Z|, Z \in \{X,Y\}$, is $O(nk)$. Furthermore, when $k \asymp n$, the total number of edges satisfies $|G^Z| \le c_0 n(n-1)$ for some constant $c_0 < \frac{1}{2}$. For weighted graphs in Table \ref{tab: para relation2}, we consider those discussed in Examples 4 and 5 ($G^X, G^Y$ can come from $k$-nearest neighbor graph, $k$-minimal spanning tree, robust $k$-nearest neighbor graph, $k$-farthest point graph, $k$-maximal spanning tree, robust $k$-farthest point graph).
\end{remark}

{ We next give concrete graph calculations for the general conditions above. {Propositions~\ref{prop:table2,verify} and~\ref{prop:table3,verify} cover both the ordinary $k$-NNG-based constructions and their robust $k$-NNG counterparts in fixed dimension, for which the needed degree and row-sum bounds can be controlled analytically.} { Supplement S9 reports numerical diagnostics for the remaining conditions for the full GIT-$\text{R}_{\text{r}k\text{NF}}$ construction, which also uses the robust $k$-farthest point graph and requires cross-component control.}}

\begin{proposition}
\label{prop:table2,verify}
Suppose that the dimension is fixed, {that Euclidean distance is used and that all pairwise distances within each sample are almost surely distinct}, and that $T_s$ is constructed using {the unweighted adjacency matrix in Example 1 based on either the $k$-NNG or the robust $k$-NNG defined in Example 5}, {with} $k \asymp n^{\alpha}$, $\alpha \in (0,1]$, and, when $\alpha=1$, $k \leq c_0 n$ for some constant $c_0 < \frac{1}{2}$. Then the constraints in Condition 2 are automatically satisfied when $\alpha \in (0,\frac{1}{2})$. Moreover, if we additionally assume %
\begin{align}
\label{eqn:A3assume,e1}
\sum_{i=1}^n |G_i^X|^2 - \frac{4|G^X|^2}{n} \asymp \sum_{i=1}^n |G_i^X|^2, \quad \sum_{i=1}^n |G_i^Y|^2 - \frac{4|G^Y|^2}{n} \asymp \sum_{i=1}^n |G_i^Y|^2,
\end{align}
{then (1) the constraints in Condition 3 are automatically satisfied when $\alpha \in (\frac{1}{2},1]$; and (2) the constraints in Condition 4 are automatically satisfied when $\alpha=\frac{1}{2}$.}
\end{proposition}

To avoid notation confusion, let $\tilde \A, \tilde \B$ denote the weight matrix before the centering, which means $A_{ij}\!=\!\tilde A_{ij}\!-\!\frac{\sum_{i \neq j} \tilde A_{ij}}{n(n-1)}, B_{ij}\!=\!\tilde B_{ij}\!-\!\frac{\sum_{i \neq j} \tilde B_{ij}}{n(n-1)}, i\!\neq\!j$. $\tilde A_{i.}$ and $\tilde B_{i.}$ are defined similarly as in~\eqref{eqn:C123,def}.

\begin{proposition}
\label{prop:table3,verify}
Suppose that the dimension is fixed and that $T_s$ is constructed using {either the graph-induced ranks from the $k$-NNG in Example 4 or the ranks based on the robust $k$-NNG in Example 5}, with $k \asymp n^{\alpha}$, $\alpha \in (0,1]$, and, when $\alpha=1$, $k \leq c_0 n$ for some constant $c_0 < \frac{1}{2}$. Then the constraints in Condition 2 are automatically satisfied when $\alpha \in (0,\frac{1}{2})$. Moreover, if we additionally assume %
\begin{align}
\label{eqn:A3assume,e4}
\sum_{i=1}^n \tilde A_{i.}^2 - \frac{(\sum_{i=1}^n \tilde A_{i.})^2}{n} \asymp \sum_{i=1}^n \tilde A_{i.}^2, \quad \sum_{i=1}^n \tilde B_{i.}^2 - \frac{(\sum_{i=1}^n  \tilde B_{i.})^2}{n} \asymp \sum_{i=1}^n \tilde B_{i.}^2 
\end{align}
{then (1) the constraints in Condition 3 are automatically satisfied when $\alpha \in (\frac{1}{2},1]$; and (2) the constraints in Condition 4 are automatically satisfied when $\alpha=\frac{1}{2}$.}
\end{proposition}

\begin{remark}
The left hand side of \eqref{eqn:A3assume,e1} represents the variability of $|G_i^X|$ and $|G_i^Y|$. It is notable that similar assumptions as \eqref{eqn:A3assume,e1} on the variability of $|G_i^X|$ have been commonly used in existing works \citep{chen2017new, zhu2023new,  zhu2024limiting}. The left hand side of \eqref{eqn:A3assume,e4}
represents the variability of $\tilde A_{i.}$ and $\tilde B_{i.}$, which can be understood as a weighted version of the assumption~\eqref{eqn:A3assume,e1}.
\end{remark}

{ Proofs of Propositions \ref{prop:table2,verify} and \ref{prop:table3,verify} are provided in Supplement S3.} In addition to the asymptotic distribution of $T_{s}, s\in \{1,2,3,4\}$, we also consider the joint distribution of $(T_{1}, T_{2}, T_{3}, T_{4})$. { The transition from Theorem~\ref{Thm: CLT general form, one stat} to Theorem~\ref{Thm: general cov chi4} is not merely a vectorization of the univariate CLT. Theorem~\ref{Thm: CLT general form, one stat} controls the marginal moments of one statistic $T_s$, through within-component quantities such as $C_{2+}^{(ss)}$ and $C_{3+}^{(ss)}$. Joint asymptotic normality additionally requires controlling mixed moments between different statistics $T_s$ and $T_{s'}$. The conditions below therefore impose cross-interaction requirements on $A_{\ell+}^{(ss')}$ and $B_{\ell+}^{(ss')}$, $\ell \in \{2,3\}$, ensuring that the corresponding mixed covariances are either asymptotically negligible or have a stable leading order. Together with uniform positive definiteness of the correlation matrix, this yields the $\chi^2_4$ approximation.} We proceed by introducing the following Conditions  \ref{Condition: Condition 1 in gene}-\ref{Condition: Condition 3 in gene}. For the asymptotic properties in Theorem \ref{Thm: general cov chi4}, only one of these conditions needs to be satisfied. For each condition, it needs to hold simultaneously for all pairs $s, s' \!\in\! \{1,2,3,4\}$ with~$s \neq s'$.

\begin{table}[!htbp]
\centering
\caption{Some scenarios satisfying Conditions \ref{Condition: Condition 1 in CLT normal}-\ref{Condition: Condition 3 in CLT normal} for unweighted graphs as described in Remark \ref{remark: cond table}, assuming $k \asymp n^{\alpha}$, $\max_{i} |G_i^Z| \asymp n^{\beta}$ and $C_{3+} \asymp n^{\nu}$, where $|G_i^Z|$ is the degree of $i$-th node in the graph $G^Z$, $\C \in \{\A, \B\}$. We also assume $C_3 \asymp n^{\nu}$ in Conditions \ref{Condition: Condition 2 in CLT normal} and \ref{Condition: Condition 3 in CLT normal}.}
\small
{
\setlength{\tabcolsep}{4pt}       %
\renewcommand{\arraystretch}{0.9} %
\begin{tabular}{lccc}
\toprule
Condition 
& $0\leq \alpha \leq 1$  
& $\alpha \leq \beta \leq 1$ 
& $1+2\alpha \leq \nu \leq 1+\alpha + \beta$ \\
\midrule
Condition \ref{Condition: Condition 1 in CLT normal} 
& $0<\alpha<\frac{1}{2}$  
& $\beta <\frac{1+\alpha}{2}$ 
& $\nu < \frac{3}{2} + \alpha$ \\
Condition \ref{Condition: Condition 2 in CLT normal} 
& $0<\alpha \leq 1$ 
& $\beta<\frac{\nu}{2}-\frac{1}{4}$ 
& $\nu>\frac{3}{2}+\alpha$ \\
Condition \ref{Condition: Condition 3 in CLT normal} 
& $0 < \alpha \leq \frac{1}{2}$ 
& $\beta < \frac{1+\alpha}{2} =\frac{\nu}{2}-\frac{1}{4}$ 
& $\nu=\frac{3}{2}+\alpha$ \\
\bottomrule
\end{tabular}
}
\label{tab: para relation1}
\end{table}

\begin{table}[!htbp]
\centering
\caption{Some scenarios satisfying Conditions \ref{Condition: Condition 1 in CLT normal}-\ref{Condition: Condition 3 in CLT normal} for weighted graphs as described in Remark \ref{remark: cond table}, with graph-induced ranks (Example 4) or ranks based on robust graph (Example 5), assuming $k \asymp n^{\alpha}$, $C_{1+} \asymp n^{\beta}$ and $C_{3+} \asymp n^{\nu}$, $\C \in \{\A, \B\}$. We also assume $C_3 \asymp n^{\nu}$ in Conditions \ref{Condition: Condition 2 in CLT normal} and \ref{Condition: Condition 3 in CLT normal}.}
\small
{
\setlength{\tabcolsep}{4pt}
\renewcommand{\arraystretch}{0.9}
\begin{tabular}{lccc}
\toprule
Condition 
& $0\leq \alpha \leq 1$  
& $2\alpha \leq \beta \leq 1+\alpha$ 
& $1+4\alpha \leq \nu \leq 1+2\alpha + \beta$ \\
\midrule
Condition \ref{Condition: Condition 1 in CLT normal} 
& $0<\alpha<\frac{1}{2}$  
& $\beta <\frac{1+3\alpha}{2}$ 
& $\nu < \frac{3}{2} + 3\alpha$ \\
Condition \ref{Condition: Condition 2 in CLT normal} 
& $0<\alpha \leq 1$ 
& $\beta<\frac{\nu}{2}-\frac{1}{4}$ 
& $\nu>\frac{3}{2}+3\alpha$ \\
Condition \ref{Condition: Condition 3 in CLT normal} 
& $0 < \alpha \leq \frac{1}{2}$ 
& $\beta < \frac{1+3\alpha}{2} =\frac{\nu}{2}-\frac{1}{4}$ 
& $\nu=\frac{3}{2}+3\alpha$ \\
\bottomrule
\end{tabular}
}
\label{tab: para relation2}
\end{table}

\begin{condition}
\label{Condition: Condition 1 in gene}
{ For each $1\leq s<s'\leq4$, $A_{2+}^{(ss')}$ and $B_{2+}^{(ss')}$ satisfy either \textnormal{(a)} $|A_{2}^{(ss')}| \asymp A_{2+}^{(ss')}$ and $|B_{2}^{(ss')}| \asymp B_{2+}^{(ss')}$, or \textnormal{(b)} $A_{2+}^{(ss')} \prec \sqrt{A_{2+}^{(ss)} A_{2+}^{(s's')}}$ and $B_{2+}^{(ss')} \prec \sqrt{B_{2+}^{(ss)} B_{2+}^{(s's')}}$. In addition, $C_{0+}^{(s)}$, $C_{1+}^{(s)}$, $C_{2+}^{(ss)}$, and $C_{3+}^{(ss)}$ satisfy Condition \ref{Condition: Condition 1 in CLT normal}, with $\mathbf{C}^{(s)} \in \{\mathbf{A}^{(s)}, \mathbf{B}^{(s)}\}$.}
\end{condition}
\begin{condition}
\label{Condition: Condition 2 in gene}
{ For each $1\leq s<s'\leq4$, $A_{3+}^{(ss')}$ and $B_{3+}^{(ss')}$ satisfy either \textnormal{(a)} $|A_{3}^{(ss')}| \asymp A_{3+}^{(ss')}$ and $|B_{3}^{(ss')}| \asymp B_{3+}^{(ss')}$, or \textnormal{(b)} $A_{3+}^{(ss')} \prec \sqrt{A_{3+}^{(ss)} A_{3+}^{(s's')}}$ and $B_{3+}^{(ss')} \prec \sqrt{B_{3+}^{(ss)} B_{3+}^{(s's')}}$. In addition, $C_{0+}^{(s)}$, $C_{1+}^{(s)}$, $C_{2+}^{(ss)}$, and $C_{3+}^{(ss)}$ satisfy Condition \ref{Condition: Condition 2 in CLT normal}, with $\mathbf{C}^{(s)} \in \{\mathbf{A}^{(s)}, \mathbf{B}^{(s)}\}$.}
\end{condition}

\begin{condition}
\label{Condition: Condition 3 in gene}
{ For each $1\leq s<s'\leq4$, $A_{2+}^{(ss')}$, $B_{2+}^{(ss')}$, $A_{3+}^{(ss')}$, and $B_{3+}^{(ss')}$ satisfy either \textnormal{(a)} $|A_{k}^{(ss')}| \asymp A_{k+}^{(ss')}$ and $|B_{k}^{(ss')}| \asymp B_{k+}^{(ss')}$ for both $k\in\{2,3\}$, or \textnormal{(b)} $A_{k+}^{(ss')} \prec \sqrt{A_{k+}^{(ss)} A_{k+}^{(s's')}}$ and $B_{k+}^{(ss')} \prec \sqrt{B_{k+}^{(ss)} B_{k+}^{(s's')}}$ for both $k\in\{2,3\}$. In addition, $C_{0+}^{(s)}$, $C_{1+}^{(s)}$, $C_{2+}^{(ss)}$, and $C_{3+}^{(ss)}$ satisfy Condition \ref{Condition: Condition 3 in CLT normal}, with $\mathbf{C}^{(s)} \in \{\mathbf{A}^{(s)}, \mathbf{B}^{(s)}\}$.}
\end{condition}
Let $\bR_n$ denote the correlation matrix of $(T_1,T_2,T_3,T_4)^{\top}$ {under the permutation null distribution}.

\begin{theorem}
\label{Thm: general cov chi4}
Assume that $\A^{(s)}$ and $\B^{(s)}$ for $s \in \{1,2,3,4\}$ satisfy Condition \ref{cond0}. If $\A^{(s)}$ and $\B^{(s)}$ further satisfy one of Conditions \ref{Condition: Condition 1 in gene}-\ref{Condition: Condition 3 in gene}, and if $\liminf_{n\rightarrow\infty}\lambda_{\min}(\bR_n)>0$, then $\Cov\big((T_1,T_2,T_3,T_4)^{\top}\big)$ is invertible for all sufficiently large $n$. {Under Condition \ref{cond0}, $\mu_s=0$ for every $s\in\{1,2,3,4\}$. Consequently,}
\begin{align}
\label{equa: chisquare converge}
(T_1, T_2, T_3, T_4 ) \Cov^{-1}\big((T_1, T_2, T_3, T_4 )^{\top}\big) (T_1, T_2, T_3, T_4 )^{\top} \stackrel{\mathcal{D}}{\rightarrow} \chi_4^2,
\end{align}
under the permutation null distribution.
\end{theorem}

The proof of Theorem \ref{Thm: general cov chi4} is provided in Supplement S4, and discussions about the invertibility of $\Cov\big((T_1, T_2, T_3, T_4 )^{\top}\big)$ are detailed in Theorem \ref{thm: cov invertible}. All of Conditions \ref{Condition: Condition 1 in gene}-\ref{Condition: Condition 3 in gene} imply that the correlation can be $O(1)$ or $o(1)$ for $T_s, T_{s'}$ with $s \neq s'$. Take Condition \ref{Condition: Condition 1 in gene} as an example. The correlation $\Cov(T_s, T_{s'})/\sqrt{\Var(T_s)\Var(T_{s'})}$ is either $o(1)$ or dominated by 

$$
A_2^{(ss')}B_2^{(ss')}/\sqrt{A_2^{(ss)}B_2^{(ss)} A_2^{(s's')}B_2^{(s's')}},
$$
which may be of order $O(1)$ when $|A_{2}^{(ss')}| \asymp A_{2+}^{(ss')}$, $|B_{2}^{(ss')}| \asymp B_{2+}^{(ss')}$, but will be of order $o(1)$ when $|A_{2}^{(ss')}| \prec A_{2+}^{(ss')}$, $|B_{2}^{(ss')}| \prec  B_{2+}^{(ss')}$.

Regarding the joint normality of multiple double-indexed permutation statistics, to the best of our knowledge, only \cite{daniels1944relation} has investigated the asymptotic behavior of bivariate double-indexed permutation statistics. In \cite{daniels1944relation}, it is required that $|A_3^{(ss')}|$ be of order $n^3 A_{0+}^{(s)} A_{0+}^{(s')}$ and $ |B_3^{(ss')}|$ be of order $n^3 B_{0+}^{(s)} B_{0+}^{(s')}$ for $s\leq s', s,s' \in \{1,2\}$. These are much stronger conditions compared to Condition \ref{Condition: Condition 2 in gene}. To see this, note that it can be shown that $|C_3^{(ss')}| \leq C_{3+}^{(ss')} \leq n^3 C_{0+}^{(s)} C_{0+}^{(s')}$, where $\C^{(s)} \in \{\A^{(s)}, \B^{(s)}\}$ and $s$ and $s'$ can be the same. Thus, $|C_3^{(ss')}| \asymp n^3C_{0+}^{(s)} C_{0+}^{(s')}$ implies $|C_3^{(ss')}| \asymp C_{3+}^{(ss')} \asymp n^3C_{0+}^{(s)} C_{0+}^{(s')}$. By similar reasoning, we have $A_{3+}^{(ss)} B_{3+}^{(ss)} \asymp n^6 (A_{0+}^{(s)} B_{0+}^{(s)})^2$, and it is straightforward to show that $n^6 (A_{0+}^{(s)} B_{0+}^{(s)})^2$ exceeds the order of $\max\{ n^{3} (A_{0+}^{(s)} B_{0+}^{(s)})^2,$ $ n (A_{1+}^{(s)} B_{1+}^{(s)})^2,$ $n A_{2+}^{(ss)} B_{2+}^{(ss)} \}$. Therefore, Condition \ref{Condition: Condition 2 in gene} is much more relaxed than the conditions in \cite{daniels1944relation}. Moreover, in our case, satisfying any one of Conditions \ref{Condition: Condition 1 in gene}, \ref{Condition: Condition 2 in gene}, or \ref{Condition: Condition 3 in gene} is sufficient. Additionally, \cite{daniels1944relation} restricts $\alpha=1$, whereas our conditions allow for $\alpha \in (0,1]$. Furthermore, \cite{daniels1944relation} restricts $\Corr(T_1, T_2)$ to be of order $1$, while our conditions permit $\Corr(T_1, T_2)$ to be either of order $1$ or $o(1)$. Overall, our framework offers greater generality for joint normality compared to \cite{daniels1944relation}.

{
We call $\Cov\big((T_1,T_2,T_3,T_4)^{\top}\big)$ asymptotically invertible if
$\liminf_{n\rightarrow\infty}\lambda_{\min}(\bR_n)>0$. More generally, given a fixed number $q$ of sequences of matrices or vectors
$\{\H_n^{(s)}\}_{n\geq 1}$, $s\in\{1,\ldots,q\}$, with $\|\H_n^{(s)}\|=1$, where $\|\cdot\|$ denotes  Frobenius norm for matrices and  Euclidean norm for vectors, we call them asymptotically linearly independent if
$
\liminf_{n\rightarrow\infty}
\inf_{\substack{\mathbf a\in\R^q\\ \|\mathbf a\|_2=1}}
\big\|\sum_{s=1}^q a_s\H_n^{(s)}\big\|^2>0.
$
We next relate asymptotic invertibility to asymptotic linear independence.

\begin{theorem}
\label{thm: cov invertible}
For each of the three cases below, the stated conditions ensure that the normalizing quantities used in that case are positive for all sufficiently large $n$. In each case,
$\Cov\big((T_1, T_2, T_3, T_4 )^{\top}\big)$ is asymptotically invertible if and only if the matrices specified in that item are asymptotically linearly independent:

\begin{itemize}
\item Under Condition \ref{cond0} and Condition \ref{Condition: Condition 1 in gene}, the four matrices $\tilde A_2^{(1)} (\tilde B_2^{(1)})^{\top}$, $\tilde A_2^{(2)} (\tilde B_2^{(2)})^{\top}$, $\tilde A_2^{(3)} (\tilde B_2^{(3)})^{\top}$, $\tilde A_2^{(4)} (\tilde B_2^{(4)})^{\top}$ are asymptotically linearly independent, where
$$
\tilde C_2^{(s)} = \big(C_{2}^{(ss)}\big)^{-\frac{1}{2}} (C_{11}^{(s)}, C_{12}^{(s)}, \ldots, C_{nn}^{(s)}) \in \R^{n^2},
$$
for $\C^{(s)} \in \{\A^{(s)}, \B^{(s)} \}, s \in \{1,2,3,4\}$.

\item Under Condition \ref{cond0} and Condition \ref{Condition: Condition 2 in gene}, the four matrices $\tilde  A_{3}^{(1)} (\tilde  B_{3}^{(1)})^{\top}$, $\tilde A_{3}^{(2)} (\tilde  B_{3}^{(2)})^{\top}$, $\tilde A_{3}^{(3)} (\tilde  B_{3}^{(3)})^{\top}$, $\tilde  A_{3}^{(4)} (\tilde  B_{3}^{(4)})^{\top}$ are asymptotically linearly independent, where
{
\begin{align}
\tilde C_3^{(s)} = \big(C_{3}^{(ss)}\big)^{-\frac{1}{2}} (C_{1.}^{(s)}, \ldots, C_{n.}^{(s)}) \in \R^n,
\label{Equa: tilde A3}
\end{align}
}
for $\C^{(s)} \in \{\A^{(s)}, \B^{(s)} \}, s \in \{1,2,3,4\}$.

\item Under Condition \ref{cond0} and Condition \ref{Condition: Condition 3 in gene}, the four matrices $\H_{23}^{(1)}, \H_{23}^{(2)}, \H_{23}^{(3)}, \H_{23}^{(4)}$ are asymptotically linearly independent, where
\begin{align*}
d_s
=(n-1)(n-2)A_2^{(ss)}B_2^{(ss)}
  +2(n+1)A_3^{(ss)}B_3^{(ss)},
\end{align*}
\vspace{-2em}
\begin{align*}
\H_{23}^{(s)}=\operatorname{diag}\Big[
\Big\{\frac{(n-1)(n-2)A_2^{(ss)}B_2^{(ss)}}{d_s}\Big\}^{1/2}
\tilde A_2^{(s)}(\tilde B_2^{(s)})^{\top},
\Big\{\frac{2(n+1)A_3^{(ss)}B_3^{(ss)}}{d_s}\Big\}^{1/2}
\tilde A_3^{(s)}(\tilde B_3^{(s)})^{\top}
\Big].
\end{align*}
\end{itemize}
\end{theorem}

The proof of Theorem \ref{thm: cov invertible} is provided in Supplement S5. Additionally, we present two corollaries that consider the $A$-side and $B$-side conditions separately, with their proofs also included in Supplement S5.

\begin{coro}
\label{coro: four AB invert}
$\Cov\big((T_1, T_2, T_3, T_4 )^{\top}\big)$ is asymptotically invertible under any one of the following sufficient conditions:
\begin{itemize}
    \item Under Condition \ref{cond0} and Condition \ref{Condition: Condition 1 in gene}, either the four vectors $\tilde A_{2}^{(1)}, \tilde  A_{2}^{(2)}, \tilde  A_{2}^{(3)}, \tilde  A_{2}^{(4)}$ or the four vectors $\tilde  B_{2}^{(1)},$ $\tilde B_{2}^{(2)},$ $\tilde  B_{2}^{(3)},$ $\tilde B_{2}^{(4)}$ are asymptotically linearly independent.
    \item Under Condition \ref{cond0} and Condition \ref{Condition: Condition 2 in gene}, either the four vectors $\tilde  A_{3}^{(1)}, \tilde  A_{3}^{(2)}, \tilde  A_{3}^{(3)}, \tilde  A_{3}^{(4)}$ or the four vectors $\tilde  B_{3}^{(1)},$ $\tilde B_{3}^{(2)},$ $\tilde  B_{3}^{(3)},$ $\tilde B_{3}^{(4)}$ are asymptotically linearly independent.
    \item Under Condition \ref{cond0} and Condition \ref{Condition: Condition 3 in gene}, at least one of the following four collections is asymptotically linearly independent: the four vectors $\tilde A_{2}^{(1)}, \tilde A_{2}^{(2)}, \tilde A_{2}^{(3)}, \tilde A_{2}^{(4)}$; the four vectors $\tilde B_{2}^{(1)}, \tilde B_{2}^{(2)}, \tilde B_{2}^{(3)}, \tilde B_{2}^{(4)}$; the four vectors $\tilde A_{3}^{(1)}, \tilde A_{3}^{(2)}, \tilde A_{3}^{(3)}, \tilde A_{3}^{(4)}$; or the four vectors $\tilde B_{3}^{(1)}, \tilde B_{3}^{(2)}, \tilde B_{3}^{(3)}, \tilde B_{3}^{(4)}$.
\end{itemize}
\end{coro}

\begin{coro}
\label{coro: two AB invert}
In our test statistic \eqref{eqn: test stat2}, with $\A^{(1)}=\A^{(2)}$, $\A^{(3)}=\A^{(4)}$, $\B^{(1)}=\B^{(3)}$, $\B^{(2)}=\B^{(4)}$, the following results hold:
\begin{itemize}
    \item Under Condition \ref{cond0} and Condition \ref{Condition: Condition 1 in gene}, $\Cov\big((T_1, T_2, T_3, T_4 )^{\top}\big)$ is asymptotically invertible if and only if $\tilde A_{2}^{(1)}$ and $\tilde A_{2}^{(3)}$ are asymptotically linearly independent, and $\tilde B_{2}^{(1)}$ and $\tilde B_{2}^{(2)}$ are asymptotically linearly independent.
    \item Under Condition \ref{cond0} and Condition \ref{Condition: Condition 2 in gene}, $\Cov\big((T_1, T_2, T_3, T_4 )^{\top}\big)$ is asymptotically invertible if and only if $\tilde A_{3}^{(1)}$ and $\tilde A_{3}^{(3)}$ are asymptotically linearly independent, and $\tilde B_{3}^{(1)}$ and $\tilde B_{3}^{(2)}$ are asymptotically linearly independent.
    \item Under Condition \ref{cond0} and Condition \ref{Condition: Condition 3 in gene}, $\Cov\big((T_1, T_2, T_3, T_4 )^{\top}\big)$ is asymptotically invertible if $\tilde A_{2}^{(1)}$ and $\tilde A_{2}^{(3)}$ are asymptotically linearly independent, and $\tilde B_{2}^{(1)}$ and $\tilde B_{2}^{(2)}$ are asymptotically linearly independent.
\end{itemize}
\end{coro}
}

Next, we evaluate the analytic $p$-value approximation based on the limiting distribution in Theorem \ref{Thm: general cov chi4}. Specifically, we examine the empirical size of GIT-$\text{R}_{\text{r}k\text{NF}}$ when $X$ and $Y$ are independent. The empirical size is calculated from $500$ replications, with $p$-values computed analytically using Theorem \ref{Thm: general cov chi4}. We consider dimensions $p \in \{20, 100, 400, 1000 \}$ and sample sizes $n \in \{50, 100, 150\}$, for $i \in [n], j \in [p]$ in the following settings: (1) Setting 0.1: $X_{ij}, Y_{ij} \overset{iid}{\sim} N(0,1)$; (2) Setting 0.2: $X_{ij}, Y_{ij} \overset{iid}{\sim} t_{10}$; (3) Setting 0.3: $X_{ij}, Y_{ij} \overset{iid}{\sim} \exp\big(N(0,1)\big)$. The empirical sizes for Settings 0.1 to 0.3 are presented in Table \ref{tab: simu, type I error}. The results show that the empirical size of GIT-$\text{R}_{\text{r}k\text{NF}}$ is reasonably well controlled across different distributions and dimensions.

\begin{table}[!htbp]
\centering
\caption{Empirical sizes of GIT-$\text{R}_{\text{r}k\text{NF}}$ at 0.05 significance level for Settings 0.1, 0.2, 0.3.}
\small                         
{
\setlength{\tabcolsep}{3pt}    
\renewcommand{\arraystretch}{0.9} 
\begin{tabular}{c|ccc|ccc|ccc}
\hline\hline
\multirow{2}{*}{$p \backslash n$} & \multicolumn{3}{c|}{Setting 0.1} 
                                  & \multicolumn{3}{c|}{Setting 0.2} 
                                  & \multicolumn{3}{c}{Setting 0.3} \\ \cline{2-10}
 & 50 & 100 & 150 & 50 & 100 & 150 & 50 & 100 & 150 \\ \hline
20   & 0.046 & 0.058 & 0.056 & 0.032 & 0.054 & 0.060 & 0.054 & 0.052 & 0.052 \\ \hline
100  & 0.050 & 0.054 & 0.046 & 0.042 & 0.046 & 0.052 & 0.046 & 0.038 & 0.046 \\ \hline
400  & 0.030 & 0.050 & 0.066 & 0.068 & 0.044 & 0.044 & 0.058 & 0.056 & 0.042 \\ \hline
1000 & 0.066 & 0.048 & 0.052 & 0.042 & 0.040 & 0.058 & 0.056 & 0.052 & 0.046 \\ \hline\hline
\end{tabular}
}
\label{tab: simu, type I error}
\end{table}

{
\subsection{Power consistency}
{ Throughout this subsection, we consider the analytically calibrated GIT that rejects the null hypothesis when $T>\chi^2_{4,1-\alpha}$, where $\chi^2_{4,1-\alpha}$ denotes the $(1-\alpha)$-quantile of the $\chi^2_4$ distribution. Under the null conditions of Theorem~\ref{Thm: general cov chi4}, this test has asymptotic level $\alpha$. Since $\chi^2_{4,1-\alpha}$ is fixed, $T\xrightarrow{\mathbb P}\infty$ implies $\mathbb P\{T>\chi^2_{4,1-\alpha}\}\to1$. We establish this power consistency for the $k$-NNG and $k$-farthest point graph constructions in Examples~1 and~4, where the graph-level sample functionals can be controlled directly, and for ranks based on robust graphs in Example~5 under a replacement-stability condition. The corresponding results are given in Theorems~\ref{thm:powerconsis,e1}--\ref{thm:powerconsis,e5}.} Because $D^Z_{ij}$ and $S^Z_{ij}$ are graph-level sample functionals (that is, they depend on the entire sample rather than on the pair $(Z_i, Z_j)$ alone), we define the following $n$-indexed population quantities:
\begin{equation}\label{eqn:thetas,e1}
\begin{aligned}
\theta_{1}^{(n)}&=\operatorname{Cov}_0\!\big(D(X_1,X_2),D(Y_1,Y_2)\big),\quad
\theta_{2}^{(n)}=\operatorname{Cov}_0\!\big(D(X_1,X_2),S(Y_1,Y_2)\big),\\
\theta_{3}^{(n)}&=\operatorname{Cov}_0\!\big(S(X_1,X_2),D(Y_1,Y_2)\big),\quad
\theta_{4}^{(n)}=\operatorname{Cov}_0\!\big(S(X_1,X_2),S(Y_1,Y_2)\big),
\end{aligned}
\end{equation}
where $\operatorname{Cov}_0$ denotes covariance under the population distribution.

\begin{theorem}[Power consistency: Example 1]\label{thm:powerconsis,e1}
{Suppose that the matrices $\mathbf S^Z$ and $\mathbf D^Z$, $Z\in\{X,Y\}$, are constructed as in Example~1 using the $k$-NNG or the $k$-farthest point graph.} If $\sqrt{n}\,|\theta_{s}^{(n)}|\to\infty$ for some $s\in\{1,2,3,4\}$, then
\[
\frac{|T_s-\mu_s|}{\sqrt{\operatorname{Var}(T_s)}}\ \xrightarrow{\mathbb{P}}\ \infty.
\]
{ If $\boldsymbol{\Sigma}$ is invertible, then $T\xrightarrow{\mathbb P}\infty$, and the rejection probability of the analytically calibrated GIT converges to one for every fixed $\alpha\in(0,1)$.}
\end{theorem}

\begin{theorem}[Power consistency: Example 4]\label{thm:powerconsis,e4}
{Suppose that the matrices $\mathbf S^Z$ and $\mathbf D^Z$, $Z\in\{X,Y\}$, are constructed as in Example~4 using graph-induced ranks from the $k$-NNG or the $k$-farthest point graph.} If $\sqrt{n}\,|\theta_{s}^{(n)}|/k^{2}\to\infty$ for some $s\in\{1,2,3,4\}$, then
\[
\frac{|T_s-\mu_s|}{\sqrt{\operatorname{Var}(T_s)}}\ \xrightarrow{\mathbb{P}}\ \infty.
\]
{ If $\boldsymbol{\Sigma}$ is invertible, then $T\xrightarrow{\mathbb P}\infty$, and the rejection probability of the analytically calibrated GIT converges to one for every fixed $\alpha\in(0,1)$.}
\end{theorem}

{

\begin{theorem}[Power consistency: Example 5]\label{thm:powerconsis,e5}
Suppose that the matrices $\mathbf S^Z$ and $\mathbf D^Z$, $Z\in\{X,Y\}$, are constructed as in Example~5. For $t\in\{1,\ldots,n\}$, let $\mathbf S^{Z,(t)}$ and $\mathbf D^{Z,(t)}$ denote the symmetrized rank matrices reconstructed after replacing $Z_t$ by an arbitrary value while keeping the remaining observations fixed. Suppose that there exist a positive sequence $\rho_n$ and a constant $C>0$, independent of $n$ and the data dimension, such that, for all $Z\in\{X,Y\}$, $t\in\{1,\ldots,n\}$, and all such replacements,
\[
\sum_{i=1}^n\sum_{j\ne i}\left|S^Z_{ij}-S^{Z,(t)}_{ij}\right|\leq C \rho_n,
\qquad
\sum_{i=1}^n\sum_{j\ne i}\left|D^Z_{ij}-D^{Z,(t)}_{ij}\right|\leq C \rho_n.
\]
If $n^{3/2}|\theta_s^{(n)}|/\{k \max(\rho_n,  k^2)\}\to\infty$ for some $s\in\{1,2,3,4\}$, then
\[
\frac{|T_s-\mu_s|}{\sqrt{\operatorname{Var}(T_s)}}\ \xrightarrow{\mathbb{P}}\ \infty.
\]
{ If $\boldsymbol{\Sigma}$ is invertible, then $T\xrightarrow{\mathbb P}\infty$, and the rejection probability of the analytically calibrated GIT converges to one for every fixed $\alpha\in(0,1)$.}
\end{theorem}
}

\noindent The proofs of Theorems \ref{thm:powerconsis,e1}--\ref{thm:powerconsis,e5} are in Supplement S10.
\begin{remark}
It is worth noting that the conditions in Theorems~\ref{thm:powerconsis,e1}--\ref{thm:powerconsis,e5} do not impose explicit restrictions on the data dimension, thereby allowing the dimension to diverge as $n$ increases. They also extend beyond conventional covariance-based dependence, since each $\theta_s^{(n)}$ is defined through the graph connection structures, allowing the tests to capture a broad range of nonlinear or higher-order dependencies between $X$ and~$Y$. Moreover, each entry $A_{ij}$ depends not only on $(X_i,X_j)$ but on the entire sample $\{X_i\}_{i=1}^n$, precluding a direct application of classical U-statistic arguments that underpin the consistency of kernel-based and distance-based tests \citep{gretton2007kernel, szekely2007measuring, rindt2021consistency}. We prove the consistency of our test statistic with a different approach.
\end{remark}
}

\section{Performance analysis}
\label{Section: Simulation studies}

In this section, we compare the performance of GIT-$\text{R}_{\text{r}k\text{NF}}$ with seven other methods: FR1, FR2, dCov, HHG, pCov, IPR, and RdCov, as introduced in Section \ref{Section: Introduction}. Here, MINT is excluded because its available implementation frequently failed at the higher dimensions considered, whereas Hallin and CPC are excluded because their computational costs were prohibitive.
Our simulation settings encompass a range of distributions to provide a comprehensive evaluation across potential real-world scenarios. In each setting, we generate $X$ and $Y$ samples with their components from (1) Gaussian, (2) $t_{10}$, and (3) log-normal distributions, covering light- to heavy-tailed and symmetric to asymmetric characteristics across low, moderate, and high-dimensional data.

Specifically, we set $\mathcal{X} = \mathcal{Y} = \R^p$ with $p \in \{20, 100, 400, 1000 \}$ and { $n \in \{50,  100,  150\}$}. We evaluate the power of each method through Settings 1.1 to 3.3, which are constructed under the alternative hypothesis. The nominal level is set as $0.05$. The power is assessed on $1000$ replications. The $p$-values for GIT-$\text{R}_{\text{r}k\text{NF}}$ are analytically approximated using Theorem \ref{Thm: general cov chi4}, while the $p$-values for the other methods are approximated through $400$ random permutations. Signal and noise levels are chosen so that the best method under each setting has moderate power. In these settings, all observations in $Y$ have one fixed dependency relationship with $X$, except in Settings 2.1 to 2.3, { where each observation independently follows one of two dependency relationships with equal probability}. The setting details are provided below, and the dependency structures summarized in Table \ref{tab: depend structure}.

For $i \in [n], j,l \in [p], Z \in \{X,Y\}$, we obtain $\sigma_i, \sigma_{i}^Z, { B_i,} B_{i}^Z \in \mathbb{R}, U_{i}, U_{i}^Z, \epsilon_{i}, \epsilon_{i}^Z, L_i \in \mathbb{R}^p$ by generating $\sigma_i, \sigma_i^Z, U_{ij}, U_{ij}^Z, \epsilon_{ij}, \epsilon_{ij}^Z, { B_i,} B_i^Z, L_{ij}$ independently. { In Settings 3.1 to 3.3, we additionally generate one matrix $\mathbf{\Theta}=[\Theta_{jl}]_{j,l=1}^p \in \mathbb{R}^{p\times p}$ in each Monte Carlo replication, independently of the other quantities, and use the same matrix for all observations.} In these settings, $\epsilon_{i1}^Z=\ldots=\epsilon_{ip}^Z$.

\begin{itemize}
\item Setting 1.1: $U_{ij} \sim N(0,1), B_{i}^X, B_{i}^Y \sim \mathrm{Bernoulli}(0.5)$, $\epsilon_{ij}^X, \epsilon_{ij}^Y \sim N(0,1.5^2)$, $X_{ij}= (-1)^{B_{i}^X}$ $\log(|U_{ij}|) + \epsilon_{ij}^X, Y_{ij}= (-1)^{B_{i}^Y}\big(5-\log(|U_{ij}|)\big) + \epsilon_{ij}^Y$.

\item Setting 1.2: $U_{ij} \sim t_{10}, B_{i}^X, B_{i}^Y \sim \mathrm{Bernoulli}(0.5)$, $\epsilon_{ij}^X, \epsilon_{ij}^Y \sim 1.5 \cdot t_{10}$, $X_{ij}= (-1)^{B_{i}^X}$ $\log(|U_{ij}|) + \epsilon_{ij}^X, Y_{ij}= (-1)^{B_{i}^Y}\big(5-\log(|U_{ij}|)\big) + \epsilon_{ij}^Y$.

\item Setting 1.3: $U_{ij} \sim \exp\big(N(-2,1)\big) , B_{i}^X, B_{i}^Y \sim \mathrm{Bernoulli}(0.5)$, $\epsilon_{ij}^X, \epsilon_{ij}^Y \sim \exp\big(N(-0.5,1)\big)$, $X_{ij}= (-1)^{B_{i}^X}\log(|U_{ij}|) + \epsilon_{ij}^X, Y_{ij}= (-1)^{B_{i}^Y}\big(5-\log(|U_{ij}|)\big) + \epsilon_{ij}^Y$.
\end{itemize}

\begin{itemize}
\item { Setting 2.1: $X_{ij} \sim N(0,1), B_i \sim \mathrm{Bernoulli}(0.5)$, $\epsilon_{ij} \sim N(0,1.2^2)$, $Y_{ij}=(1-B_i)\log(|X_{ij}|)+B_i\exp(0.6X_{ij})+\epsilon_{ij}$.}

\item { Setting 2.2: $X_{ij} \sim t_{10}, B_i \sim \mathrm{Bernoulli}(0.5)$, $\epsilon_{ij} \sim t_{10}$, $Y_{ij}=(1-B_i)\log(|X_{ij}|)+B_i\exp(0.5X_{ij})+\epsilon_{ij}$.}

\item { Setting 2.3: $X_{ij} \sim \exp\big(N(-4,1)\big), B_i \sim \mathrm{Bernoulli}(0.5)$, $\epsilon_{ij} \sim \exp\big(N(-4,2.9^2)\big)$ for $p=20$, $\epsilon_{ij} \sim \exp\big(N(-4,2.3^2)\big)$ for $p=100$, $\epsilon_{ij} \sim \exp\big(N(-4,2^2)\big)$ for $p \in \{400,1000\}$, $Y_{ij}=(1-B_i)\log(X_{ij})+B_i\exp(0.7X_{ij})+\epsilon_{ij}$.}

\item Setting 3.1: $L_{ij}, \Theta_{jl} \! \sim \! N(0,1)$, $\epsilon_{i1}^X, \epsilon_{i1}^Y \! \sim \! N(0, 8^2),$ $X_i \!= \! L_i^{\top} \sin \mathbf{\Theta}+\epsilon_{i}^X$, $Y_i\!=\! L_i^{\top} \cos \mathbf{\Theta} +\epsilon_{i}^Y$.

\item Setting 3.2: $L_{ij}, \Theta_{jl} \sim t_{10}$, $\epsilon_{i1}^X, \epsilon_{i1}^Y \sim 10 \cdot t_{10},$ $X_i=L_i^{\top} \sin \mathbf{\Theta}+\epsilon_{i}^X$, $Y_i= L_i^{\top} \cos \mathbf{\Theta}+\epsilon_{i}^Y$.

\item Setting 3.3: $L_{ij}, \Theta_{jl} \sim \exp\big(N(0,1)\big)$, $\epsilon_{i1}^X, \epsilon_{i1}^Y \sim \exp(N(0, 25^2)),$ $X_i=L_i^{\top} \sin \mathbf{\Theta}+\epsilon_{i}^X$, $Y_i= L_i^{\top} \cos \mathbf{\Theta}+\epsilon_{i}^Y$.
\end{itemize}

\begin{figure}[htbp]
\centering
\includegraphics[width=0.97\linewidth]{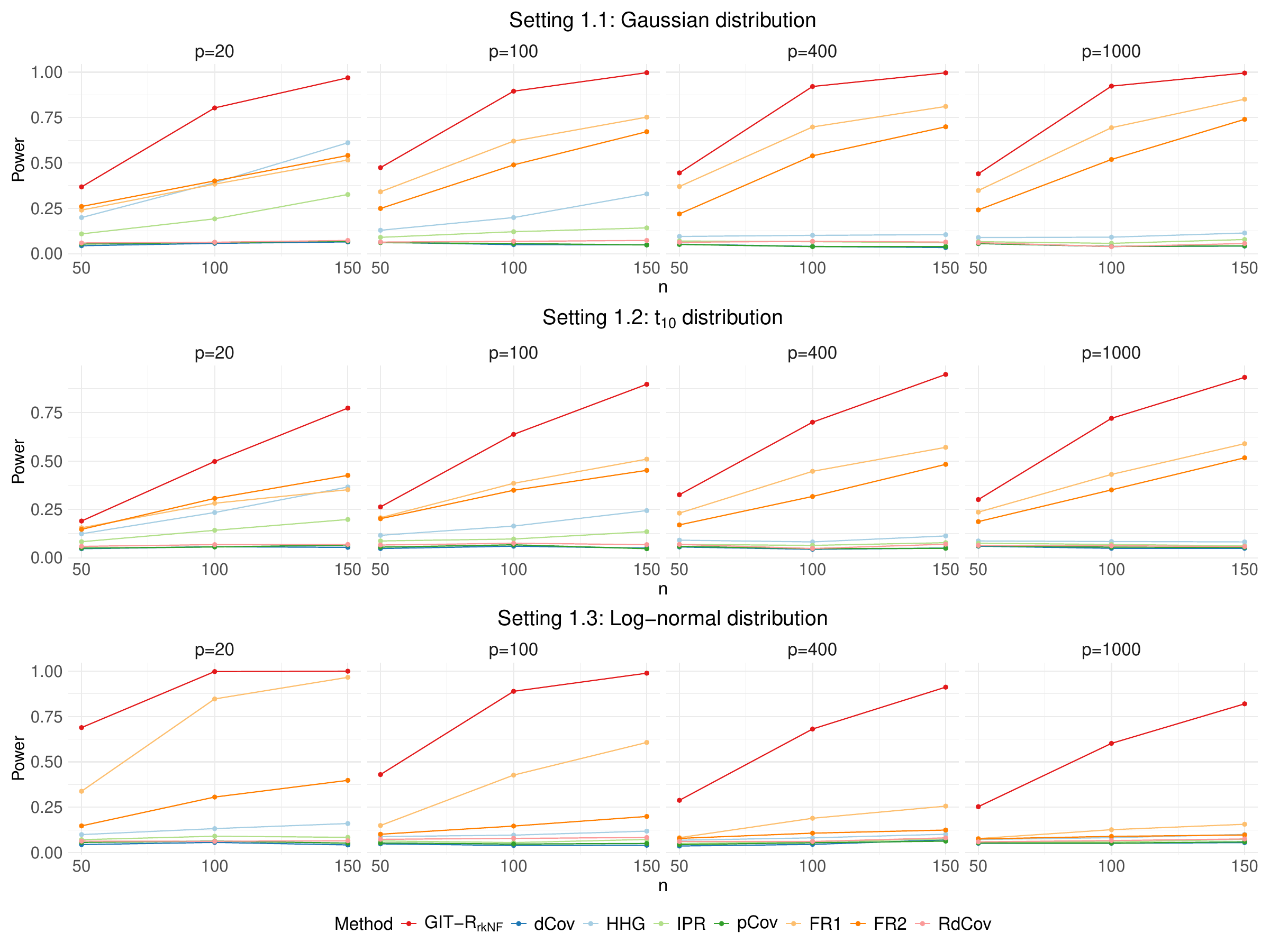}
\caption{Estimated power for Settings 1.1 to 1.3}
\label{fig:power-settings-1}
\end{figure}

\begin{figure}[htbp]
\centering
\includegraphics[width=0.97\linewidth]{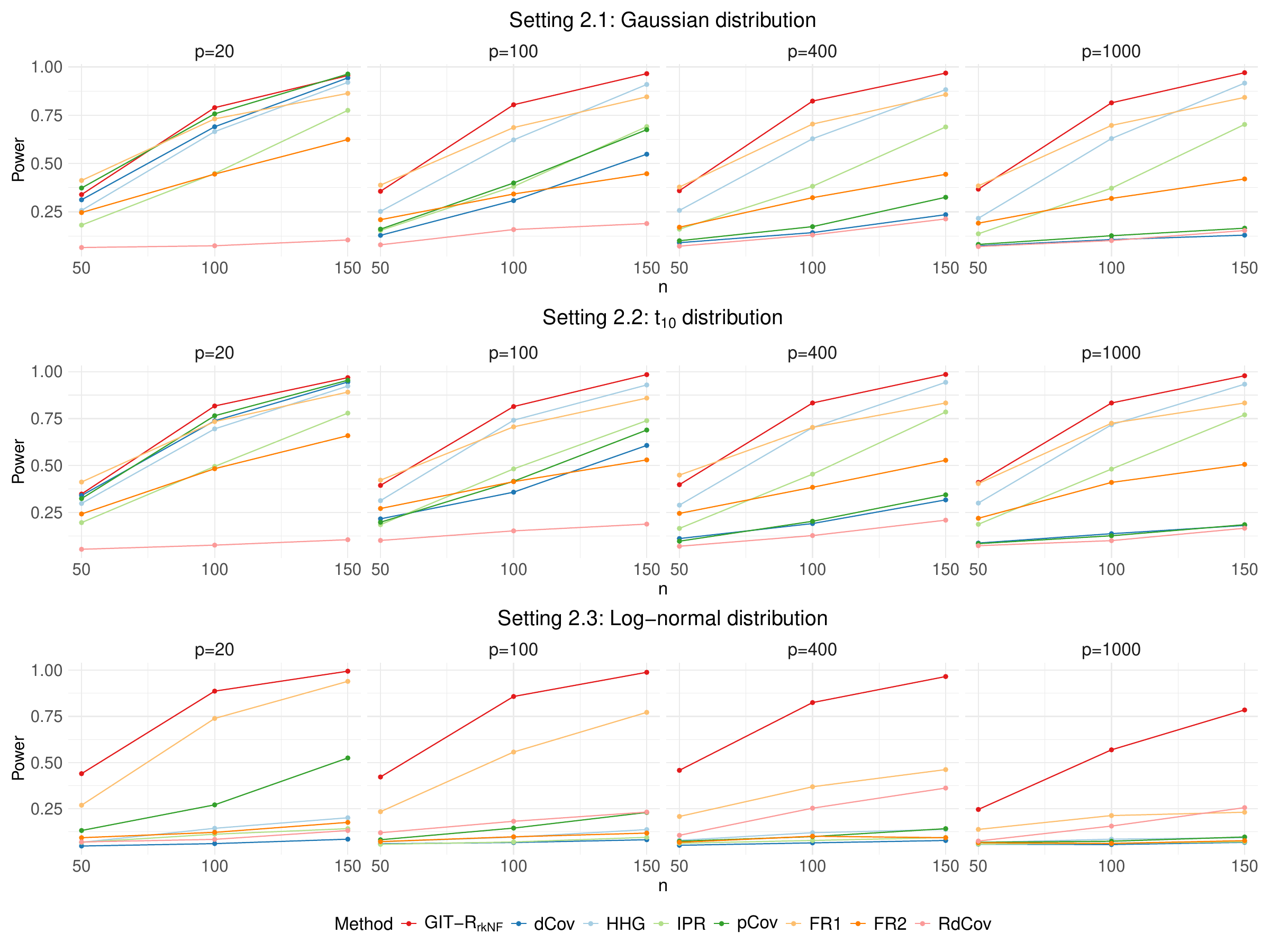}
\caption{Estimated power for Settings 2.1 to 2.3 }
\label{fig:power-settings-2}
\end{figure}

\begin{figure}[htbp]
\centering
\includegraphics[width=0.97\linewidth]{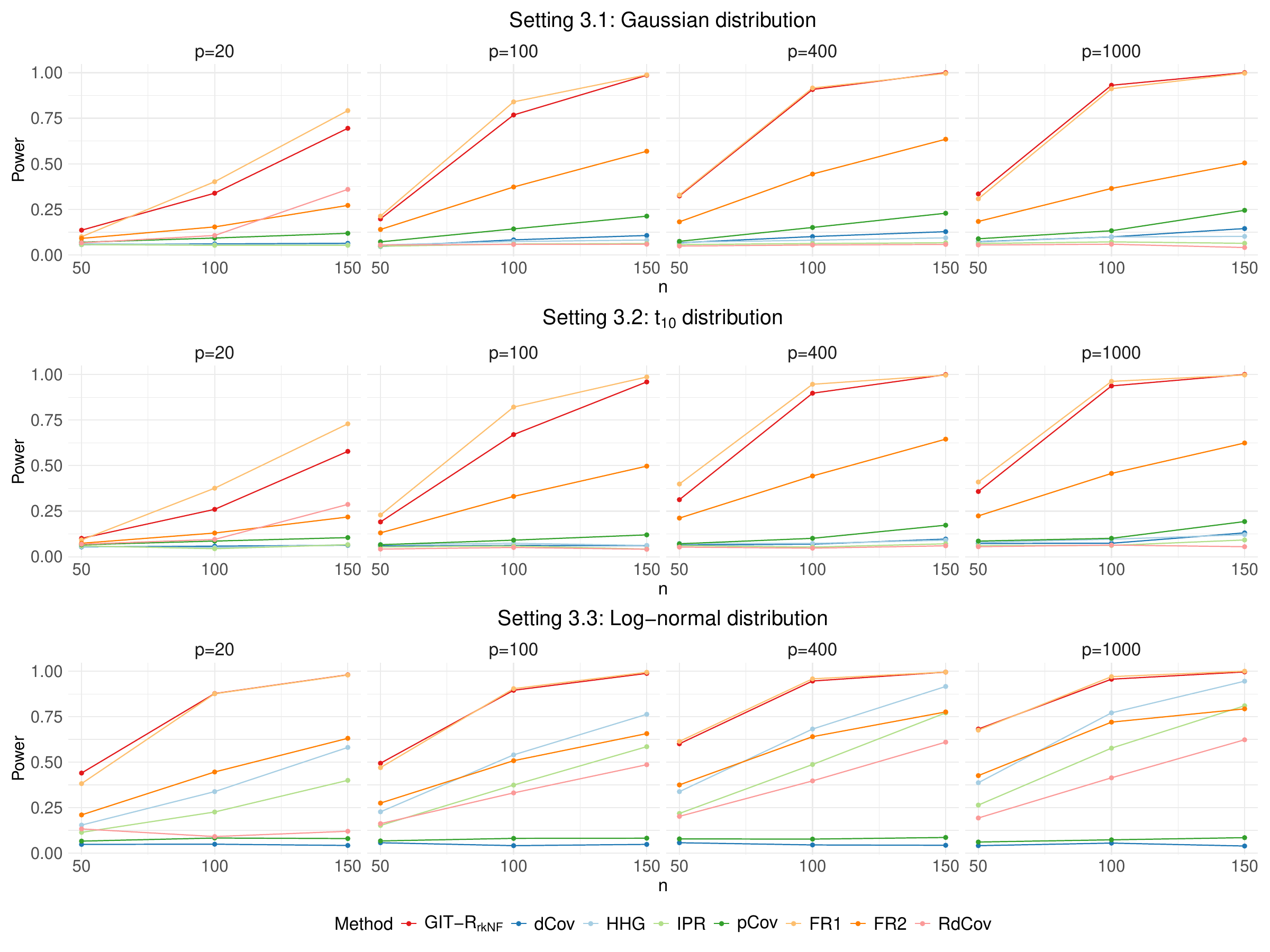}
\caption{Estimated power for Settings 3.1 to 3.3 }
\label{fig:power-settings-3}
\end{figure}

\begin{figure}[htbp]
\centering
\includegraphics[width=1\linewidth]{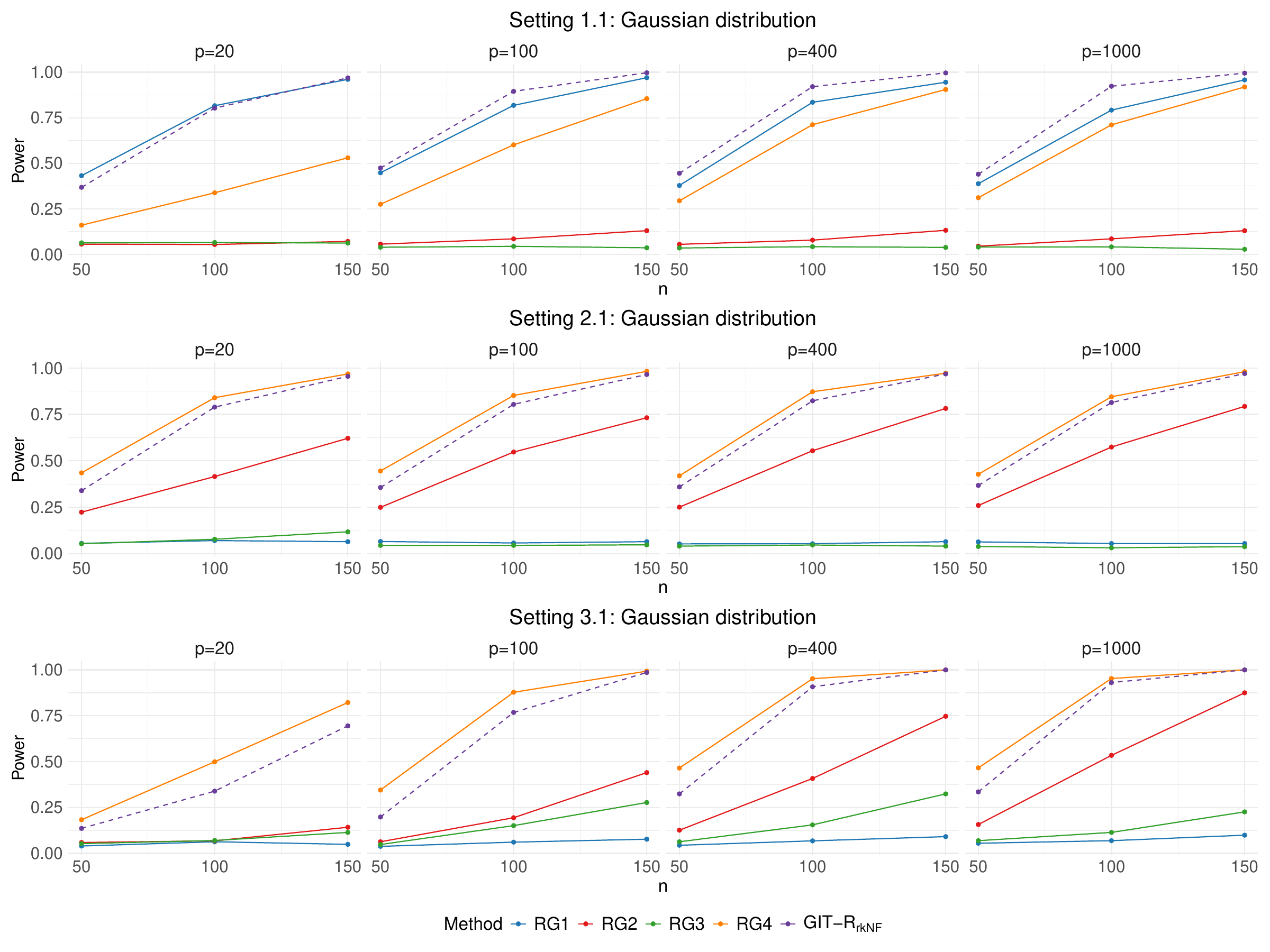}
\captionsetup{width=\textwidth}
\caption{Estimated power of components for Settings 1.1, 2.1, 3.1}
\label{fig:component-power-gaussian}
\end{figure}

We begin by analyzing the estimated power for Settings 1.1 through 3.3, as shown in Figures \ref{fig:power-settings-1} to \ref{fig:power-settings-3}. Following this, we examine the contributions of each component in GIT-$\text{R}_{\text{r}k\text{NF}}$ by evaluating the empirical power of the test and its individual components, named by RG1, RG2, RG3, and RG4, which are based on the test statistic
$
\frac{|T_s-\E(T_s)|}{\sqrt{\Var(T_s)}},  s \in \{1,2,3,4\}
$. Figure \ref{fig:component-power-gaussian} compares these components, highlighting the relative importance of each pairwise relationship for settings generated from Gaussian distributions. Similar results for the other two distributions are provided in Supplement S8.

We first examine results for Settings 1.1-1.3, as shown in Figure \ref{fig:power-settings-1}. In Setting 1.1 for the Gaussian distribution, GIT-$\text{R}_{\text{r}k\text{NF}}$ exhibits the highest power, with FR1 and FR2 also showing moderate power, while most other methods have negligible power. Notably, HHG maintains moderate power only at $p=20$. For the $t_{10}$ distribution in Setting 1.2, GIT-$\text{R}_{\text{r}k\text{NF}}$ again achieves the highest power, with FR1, and FR2 also showing adequate power. Finally, for the log-normal distribution in Setting 1.3, GIT-$\text{R}_{\text{r}k\text{NF}}$ significantly outperforms all other methods. FR1 and FR2 show moderate power at low dimensions but their power drops sharply at high dimensions.

From Figure \ref{fig:power-settings-2}, for the Gaussian distribution in Setting 2.1, most methods demonstrate~satisfactory power, except for RdCov, which has almost no power. As the dimension increases, GIT-$\text{R}_{\text{r}k\text{NF}}$ performs the best, followed closely by FR1, with HHG also showing good power. In Setting 2.2, under the $t_{10}$ distribution, most methods perform well at low dimensions. As the dimension increases, GIT-$\text{R}_{\text{r}k\text{NF}}$ performs the best, followed closely by FR1 and HHG. For the log-normal distribution in Setting 2.3, GIT-$\text{R}_{\text{r}k\text{NF}}$ exhibits outstanding power compared to all other methods across all dimensions. While FR1 and pCov perform moderately well at low~dimensions, with FR1 outperforming pCov, their gap in performance relative to GIT-$\text{R}_{\text{r}k\text{NF}}$ widens as $p$ increases.

\begin{table}[!htbp]
\centering
\caption{Dependency structures in Settings 1.1 to 3.3: Hierarchical (layered dependency structure); { Mixture (observation-level mixture of two dependency types)}; Non-monotonic (non-monotonic dependency trends).}
\small
{
\setlength{\tabcolsep}{4pt}
\renewcommand{\arraystretch}{0.9}
\begin{tabular}{lccc}
\toprule
Setting & Hierarchical & { Mixture} & Non-monotonic \\
\midrule
1.1--1.3 & $\checkmark$ &            & $\checkmark$ \\
2.1--2.3 &              & $\checkmark$ & $\checkmark$ \\
3.1--3.3 & $\checkmark$ &            & $\checkmark$ \\
\bottomrule
\end{tabular}
}
\label{tab: depend structure}
\end{table}

From Figure \ref{fig:power-settings-3}, we observe that for the Gaussian distribution in Setting 3.1, GIT-$\text{R}_{\text{r}k\text{NF}}$  and FR1 are the two leading methods. FR2 achieves about half the power of GIT-$\text{R}_{\text{r}k\text{NF}}$, while the other methods exhibit insufficient power. For the $t_{10}$ distribution in Setting 3.2, GIT-$\text{R}_{\text{r}k\text{NF}}$ and FR1 continue to lead, and are followed by FR2. All other methods exhibit negligible power. In Setting 3.3, for the log-normal distribution, GIT-$\text{R}_{\text{r}k\text{NF}}$ and FR1 outperform all other methods, followed by FR2, HHG and IPR.

We now examine the individual components of GIT-$\text{R}_{\text{r}k\text{NF}}$. As shown in Figure \ref{fig:component-power-gaussian}, RG1 and RG4 are the most influential in Setting 1.1, indicating the dependency involving the alignment of dissimilarity in $X$ with dissimilarity in $Y$, as well as similarity in $X$ with similarity in $Y$. In Setting 2.1, RG2 and RG4 are the most impactful, indicating a dependency involving the alignment of dissimilarity in $X$ with similarity in $Y$, and similarity in $X$ with similarity in $Y$. Finally, in Setting 3.1, RG4 emerges as the most influential component, highlighting a strong dependency between similarity in $X$ and similarity in $Y$.

In summary, GIT-$\text{R}_{\text{r}k\text{NF}}$ consistently demonstrates superior performance across a variety of distributions and settings, particularly under complex dependency between $X$ and $Y$. FR1 and HHG generally perform well, but their effectiveness is not stable when the dependency becomes complicated. Additionally, FR1 tends to underperform in settings involving high dimensions or log-normal distributions. Based on these comprehensive results, we recommend GIT-$\text{R}_{\text{r}k\text{NF}}$ for its overall effectiveness across diverse settings.

\section{Real data analysis}
\label{Section: Real data analysis}

{ The Genotype-Tissue Expression (GTEx) project studies the relationship between genetic variation and gene expression in humans. To illustrate the proposed tests, we analyze gene-expression data from GTExPortal (version V8)\footnote{\href{https://www.gtexportal.org/home/datasets}{https://www.gtexportal.org/home/datasets}}. Previous analyses have considered associations among GTEx expression categories \citep{urbut2019flexible, zhou2020unified, khunsriraksakul2022integrating}. This application considers the GTEx tissue category ``Breast - Mammary Tissue'' and the cell-line category ``Cells - EBV-transformed lymphocytes.''}

{ The dataset used here, obtained from \citet{khunsriraksakul2022integrating}, was adjusted for sex, sequencing platform, the top three genetic principal components, and probabilistic estimation of expression residuals (PEER) factors. Whether Epstein--Barr virus (EBV) promotes breast cancer remains unresolved, and the underlying mechanisms are not fully understood \citep{hu2016epstein, arias2022epstein}. \citet{hu2016epstein} reported that mammary epithelial cells express CD21, can be infected by EBV, and undergo changes in gene expression following EBV infection. Against this biological background, we test independence between the gene-expression profiles of the two GTEx categories using $66$ common samples. The data contain expression measurements for $25,849$ genes in ``Breast - Mammary Tissue'' and $22,759$ genes in ``Cells - EBV-transformed lymphocytes.''}

\begin{table}[!htbp]
\caption{{ Permutation $p$-values for GIT-$\text{R}_{\text{r}k\text{NF}}$ and the comparison methods, based on $1,000$ permutations.}}
\label{Table: pvalue comparison}
\centering
\begin{tabular}[t]{l|c|c|c|c|c|c|c|c}
\hline
 & GIT-$\text{R}_{\text{r}k\text{NF}}$ & dCov & HHG & { pCov} & IPR & { RdCov} & FR1 & FR2 \\
\hline
$p$-value  & $\mathbf{0.015}$ & 0.331 & 0.524 & { $\mathbf{0.033}$} & 0.593 & { 0.677} & 0.789 & 0.891\\
\hline
\end{tabular}
\end{table}

\begin{table}[h]
\caption{{ Permutation $p$-values for RG1--RG4, based on $1,000$ permutations.}}
\label{Table: RG pvalue comparison}
\centering
\begin{tabular}[!htbp]{l|c|c|c|c}
\hline
& RG1 & RG2 & RG3 & RG4 \\
\hline
$p$-value  & 0.216 & 0.320 & $\mathbf{0.004}$ & 0.698 \\
\hline
\end{tabular}
\end{table}

{ We compare GIT-$\text{R}_{\text{r}k\text{NF}}$ with the seven methods used in Section \ref{Section: Simulation studies}: FR1, FR2, dCov, HHG, pCov, IPR, and RdCov. For each method, we use $1,000$ permutations to estimate the $p$-value. As shown in Table \ref{Table: pvalue comparison}, the permutation $p$-values for GIT-$\text{R}_{\text{r}k\text{NF}}$ and pCov are $0.015$ and $0.033$, respectively, and both tests reject independence at the $0.05$ level. The remaining six methods yield $p$-values between $0.331$ and $0.891$.}

\begin{figure}[t]
\centering
\includegraphics[width=0.6\linewidth]{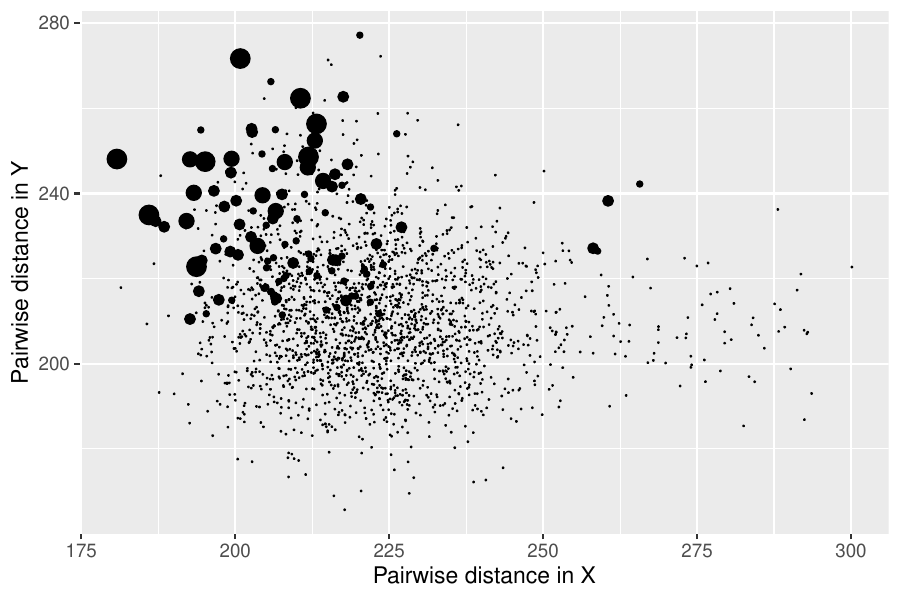}
\caption{Scatter plot of pairwise distances in $X$ and $Y$, with point size representing the magnitude of $S_{ij}^XD_{ij}^Y/25+1$, rounded to the nearest integer.}
\label{Figure: Scatter plot in real data}
\end{figure}

To further investigate contributions from different components, we analyze RG1 through RG4 to assess the pairwise distance relationships between observations of $X$ and observations of $Y$ that contribute to the significance of GIT-$\text{R}_{\text{r}k\text{NF}}$. Table \ref{Table: RG pvalue comparison} shows that among four components of GIT-$\text{R}_{\text{r}k\text{NF}}$, only RG3 shows a significant result, with the permutation $p$-value of $0.004$. Figure \ref{Figure: Scatter plot in real data} depicts the pairwise distances for samples connected in both the similarity graph of $X$ and the dissimilarity graph of $Y$. The plot suggests an inverse association between the pairwise distances in $X$ and $Y$ for these selected pairs.

{ Using the analytic approximation, the $p$-value for GIT-$\text{R}_{\text{r}k\text{NF}}$ is 0.008. The permutation calibration based on 1,000 permutations gives a $p$-value of 0.015. Both approaches therefore reject independence at the 0.05 significance level for the 66 matched samples.}

\section{Conclusion and Discussion}
\label{sec: discussion}
In summary, we propose a novel framework for testing independence between random objects that simultaneously incorporates both similarity and dissimilarity information. From a theoretical standpoint, we establish univariate and multivariate combinatorial central limit theorems for double-indexed statistics under flexible graph sparsity conditions, extending classical results to a new domain. Our analysis bridges the gap between sparse and dense graph regimes. Moreover, we show the power consistency when the test statistic comes from Examples 1 or 4, and under a replacement-stability condition when it comes from Example 5. Our test remains reliable in the presence of outliers or heavy-tailed distributions, and it can be applied~to non‐Euclidean data if appropriate similarity and dissimilarity measures are available.

{

A fundamental challenge is to establish a useful order for the replacement-stability rate $\rho_n$ in Theorem~\ref{thm:powerconsis,e5} directly for ranks based on robust graphs. The concept of robust graphs is recent \citep{zhu2023new} and the relevant probabilistic tools are still lacking. Establishing such an order from the robust graph construction therefore remains an open and intriguing problem. We hope that the present work stimulates further theoretical and methodological developments along these lines and paves the way for broader applications of {methods based on ranks from robust graphs} in modern high-dimensional data analysis.

}

\section*{Acknowledgement}

Mingshuo Liu and Hao Chen were supported in part by the U.S. National Science Foundation under Grant DMS-2311399.

\bibliographystyle{biometrika}
\bibliography{library}
\end{document}